\documentclass[10pt,aps,prd,showpacs,nofootinbib,floats,floatfix,preprintnumbers,groupedaddress,twocolumn]{revtex4-1}

\usepackage{aas_macros}

\usepackage[normalem]{ulem} 
\usepackage{bm}
\usepackage{hyperref}
\usepackage{latexsym}
\usepackage{dcolumn}
\usepackage{amsmath,amsfonts,amssymb}
\usepackage{graphicx,epsfig}
\usepackage{fancyhdr}
\usepackage{hyperref}
\usepackage{graphicx,epstopdf}
\usepackage{xcolor}

\usepackage{aas_macros}
\usepackage{xcolor}
\usepackage{graphicx,hyperref}
\usepackage{dcolumn}
\usepackage{bm}
\usepackage{ulem}
\usepackage{amsmath}
\usepackage{mathrsfs}
\usepackage{float}
\usepackage{natbib}
\usepackage{color}

\begin{document}

\title{Relativistic Accretion Flow in a Generic Class of Spherically Symmetric Static Spacetime}

\author{Pradeepkumar Yadav}\email{pradeepphy@iitg.ac.in}
\author{Sayan Chakrabarti}\email{sayan.chakrabarti@iitg.ac.in}
\author{Santabrata Das}\email{sbdas@iitg.ac.in (Corresponding Author)}

\affiliation{Indian Institute of Technology Guwahati, Guwahati, 781039, Assam, India}
\date{\today}




\begin{abstract}

    We investigate the properties of low angular momentum, inviscid, advective accretion flows in a generic static and spherically symmetric spacetime that incorporates higher-order corrections up to the fourth order in $1/r$. Employing this metric, we self-consistently solve the relativistic hydrodynamical equations and obtain the family of global transonic accretion solutions (O, A, W and I-types) by means of the spacetime parameters $(\delta, \eta, \beta)$ and the flow parameters (specific energy $\mathscr{E}$ and angular momentum $\lambda$). Our analysis reveals that the accretion flow possesses either single or multiple critical points depending on these input parameters. We delineate the regions of the $\delta-\lambda$ and $\lambda-\mathscr{E}$ parameter spaces that admits solutions with multiple critical points and demonstrate how these regions evolve with increasing spacetime parameter $\delta$. Furthermore, while connecting the spacetime geometry with observable signatures, we compute the spectral energy distribution (SED) from thermal bremsstrahlung emission and observe that increasing $\delta$ enhances the SED relative to the Schwarzschild case. Finally, we find that global transonic solutions harboring inner critical points (I-types) yields more luminous power than those with only outer critical points (O and A types).
    
\end{abstract}
 
\maketitle

\section{Introduction}\label{sec:intro}

Black holes (BHs) are among the most intriguing cosmic objects predicted by the general theory of relativity \citep{shapiro1983black}. Accretion serves as a crucial pathway for extending our understanding of these extremely compact objects in the universe \citep{frank2002accretion}. The accretion process around BHs is widely regarded as the fundamental mechanism capable of explaining the observed radiation from various astrophysical systems, including quasars \citep{shen2014diversity,dexter2010quasar,proga2007dynamics}, active galactic nuclei \citep{peterson1997introduction,fabian2012observational}, and BH X-ray binaries \citep{esin1997advection,davis2006testing}. Since BHs themselves do not emit radiation, understanding the dynamics of accreting matter is essential for inferring their physical properties. Over the past decades, numerous studies have explored BH accretion under a variety of physical conditions (\citealt{igumenshchev1999rotating,li2013rotating,yuan2014hot}, and references therein). In parallel, accretion onto several classes of exotic compact objects, such as boson stars \citep{torres2002accretion,guzman2006accretion}, wormholes \citep{harko2009thin}, gravastars \citep{harko2009can}, quark stars \citep{kovacs2009can}, and naked singularities \citep{joshi2013distinguishing,kovacs2010can}, has also been investigated. Furthermore, hydrodynamical studies of accretion have been extended to a variety of modified gravity frameworks, including Chern–Simons gravity \citep{harko2010thin}, braneworld BHs \citep{pun2008thin,heydari2010black}, Ho\v{r}ava-Lifshitz gravity \citep{harko2011thin,harko2009testing}, and $f(R)$ gravity \citep{uniyal2024}, among others.

For almost a century, Einstein’s General Relativity (GR) has successfully passed a wide range of experimental tests ranging from classical solar system experiments to the recent detections of gravitational waves \citep{will2018theory, abbott2016gw150914}. However, several theoretical and observational challenges, such as the lack of a theory of gravity consistent with quantum mechanics, the existence of dark matter and dark energy, singularities at the center of BHs and at the beginning of the big bang, have motivated the development of alternative theories of gravity \cite[]{clifton2012modified, nojiri2011unified}. In many such theoretical frameworks, the Schwarzschild metric is modified by adding inverse power law terms as correction in the metric function. For instance, coupling GR with electromagnetism leads to the Reissner-Nordstr\"{o}m solution, which introduces a $1/r^2$ term in the metric function \cite[]{nordstrom1918energy}. Higher-dimensional theories such as Gauss–Bonnet and Lovelock gravity, which appear as low energy limits of string theory, generate $1/r^4$ corrections to BH solutions \citep{cai2002gauss, lovelock1971einstein}. Similarly, Braneworld models including Randall–Sundrum scenarios induce effective $1/r^4$ terms in four-dimensional spacetime arising from a `tidal charge’ associated with gravitational leakage into extra dimensions \citep{dadhich2000black}. Quantum corrected BH solutions derived from effective field theory approaches or one-loop quantum gravity calculations also introduce higher order terms to the metric \citep{kirilin2002quantum, bjerrum2003quantum, casadio2002new, Devi:2021ctm}.

Such higher order modifications to the metric influence several key observables near compact objects including the photon sphere, the gravitational redshift, and the innermost stable circular orbit (ISCO), all of which are sensitive to the underlying spacetime geometry near compact object \cite[]{chandrasekhar1998mathematical, cardoso2019testing, bardeen1973four}. These modifications are expected to affect the structure and dynamics of accretion flows around the central objects. Since gravitational fields are strongest in the vicinity of the BHs, accreting matter approaching the event horizon provides a natural and effective probe for testing the alternative theories of gravity. Motivated by these considerations, we adopt the metric proposed by \citet{mafuz2024shadows}, which includes correction terms up to fourth order in $1/r$. Their approach provides a phenomenological  generalization of static, spherically symmetric spacetimes allowing us to test, in a model independent way, how deviations from standard Schwarzschild or Reissner-Norsdstr\"{o}m geometry would affect observable features of astrophysical compact objects. It also has the potential to offer a bridge between theory and observation: by constraining the coefficients through different observational data. In this work, using this metric, we carry out a comprehensive hydrodynamical analysis of low angular momentum accretion flow in a static, spherically symmetric BH spacetime.

Within this framework, we adopt a relativistic equation of state \cite[EoS;][]{chatto2009effects} and perform the critical point analysis to calculate and classify the critical points. We then compute the global transonic accretion solutions as functions of the spacetime parameters ($\delta$, $\eta$, and $\beta$) and the input flow parameters, namely the specific energy $\mathscr{E}$ and the angular momentum $\lambda$. This enables us, for the first time to the best of our knowledge, to examine how the variations of the spacetime parameters modify the accretion solutions. We further determine the allowed parameter ranges in the $\delta-\lambda$ and $\lambda-\mathscr{E}$ planes that give rise to multiple critical points. Finally, we evaluate the spectral energy distribution (SED) associated with the global transonic solutions and examine the role of the parameter $\delta$ in regulating the resulting SEDs.

This paper is organized as follows. Section~\ref{sec:metric} provides a brief overview of the chosen metric. In Section~\ref{sec:method}, we present the model assumptions followed by the governing equations and the critical point analysis. We present results in section \ref{sec:accretion}. Finally, Section~\ref{sec:summary} summarizes our key findings.

\section{Description of Generic Spherically Symmetric Spacetime} \label{sec:metric}

We begin with an asymptotically flat static spherically symmetric spacetime, where the line element is expressed as \cite[]{mafuz2024shadows},
\begin{align}\label{metric}
    \begin{split}
    ds^2 = & ~g_{tt}dt^2 + g_{rr}dr^2 + g_{\theta\theta}d\theta^2 + g_{\phi\phi}d\phi^2\\
    =& -\mathscr{F}(r) dt^2 + \mathscr{F}^{-1}(r) dr^{2}
    + r^2 d\theta^2 + r^2 \rm sin^2 \theta ~d\phi^2.
    \end{split}
\end{align}
In equation \eqref{metric}, $\mu$ and $\nu$ are the spacetime indices running from $0 \rightarrow 3$ and $g_{\mu \nu}$ are the components of the metric under consideration. The component $\mathscr{F}(r)$ for the generic class of spacetime is defined as,
\begin{equation} \label{metriccomp}
    \mathscr{F}(r) = \Bigg(1 + \frac{\mathcal{P}}{r} + \frac{\mathcal{Q}}{r^2} + \frac{\mathcal{S}}{r^3} 
    + \frac{\mathcal{T}}{r^4}\Bigg),
\end{equation}
where $\mathcal{P}$, $\mathcal{Q}$, $\mathcal{S}$, and $\mathcal{T}$ are constants. Note that the maximum power of the term $1/r$ is restricted to four that allows a tractable and meaningful analysis of the corresponding accretion solutions. Needless to mention that different combinations of these constants in equation (\ref{metriccomp}) yield numerous examples of metrics in GR and modified theories gravity \cite[see][and references therein]{kiselev2003quintessence, heydarzade2017black, canate2020novel, mafuz2024shadows}. Allowing multiple powers of $1/r$ with independent coefficients provides enough flexibility to capture a large variety of possible deviations from the standard GR black holes in vacuum, which may arise due to additional fields, matter distributions, modifications of gravity, or phenomenological corrections.

Here, we assume that the metric represents a solution within GR or a modified theory of gravity that satisfies the Einstein or Einstein-like field equations, i.e. $G_{\mu\nu} = 8\pi T^{\rm eff}_{\mu\nu}$, where $T^{\rm eff}_{\mu\nu}$ denotes the effective stress-energy tensor. Indeed, $T^{\rm eff}_{\mu\nu}$ encapsulates the properties of the matter distribution under consideration, while the Einstein tensor $G_{\mu\nu}$ underlines the spacetime curvature. Throughout this study, we adopt the convention that Greek indices such as $\mu$, $\nu$, $\alpha$ run over the spacetime coordinates ($t$,$r$,$\theta$,$\phi$). Following \cite{mafuz2024shadows}, we express $\mathcal{P}$ in terms of the BH mass as $-2M_{\rm BH}$ because the inequalities arising from the Weak Energy Condition (WEC) and the Null Energy Condition (NEC) are independent of the parameter $\mathcal{P}$. Furthermore, we introduce dimensionless parameters $\delta$, $\eta$, and $\beta$ as,
\begin{equation}
    \mathcal{Q} = \delta M_{\rm BH}^{2};~ \mathcal{S} = \eta M_{\rm BH}^{3}; ~\text{and}~ \mathcal{T} = \beta M_{\rm BH}^{4}.
\end{equation}
In this work, we use a unit system as $M_{\rm BH}=G=c=1$, where $M_{\rm BH}$ is the mass of the black hole, $G$ is the gravitational constant and $c$ is the speed of light. In this system, length, angular momentum and energy is expressed in units of $r_g=GM_{\rm BH}/c^2$, $cr_g$ and $c^2$, respectively. Accordingly, we rewrite equation (\ref{metriccomp}) as,
\begin{equation} \label{metriccomp_dimless}
	\mathscr{F}(r) = \left(1 - \frac{2}{r} + \frac{\delta}{r^2} + \frac{\eta}{r^3} + \frac{\beta}{r^4}\right).
\end{equation}

It is worth mentioning that when $\delta = \eta = \beta = 0$, the metric simplifies to the Schwarzschild spacetime of a non-rotating BH. The horizon condition for the generic metric is given by, 
\begin{equation}\label{horizoncondn}
    g^{rr}(r_{\rm H}) = \mathscr{F}(r) = \left(1 - \frac{2}{r} + \frac{\delta}{r^2} + \frac{\eta}{r^3} + \frac{\beta}{r^4}\right) = 0.
\end{equation}
By analytically solving equation~(\ref{horizoncondn}), we determine the location of the event horizon $r_{\rm H}$ (see Appendix~\ref{eventhorizon} for details). In this study, we exclusively focus on accretion solutions corresponding to BHs, which is eventually ensured by choosing only those combinations of the parameters $\delta$, $\eta$, and $\beta$ that satisfy the horizon condition and thus confirm the existence of the event horizon. To ensure this and to satisfy different energy conditions, in the present study, we have restricted all the deformation parameters to be positive below their respective upper bounds. Negative choices of $\delta$, $\eta$ and $\beta$ exist and different signature combinations of these parameters can give rise to violations of weak and null energy conditions (see \cite{mafuz2024shadows} for more detail).

\section{Basic Assumptions and Model Formalism}\label{sec:method}

In this work, we develop our analysis within the framework of a steady, inviscid, low angular momentum accretion flow that advects toward the BH along the equatorial plane. Such flows provide a physically motivated description of accretion in strong gravity environments \citep{chakrabarti1989standing, das2001standing, dihingia2019shocks}. We adopt a generic, static, and spherically symmetric spacetime background to capture the essential curvature effects governing the radial inflow dynamics. Within this framework, the relativistic hydrodynamic equations governing mass, momentum, and energy conservation form the basis of the accretion model and are developed in detail in the following subsection. In this work, we adopt an unit system as $M_{\rm BH}=G=c=1$, where $M_{\rm BH}$ is the mass of the black hole, $G$ is the gravitational constant and $c$ is the speed of light. In this system, length, angular momentum and energy is expressed in units of $r_g=GM_{\rm BH}/c^2$, $cr_g$ and $c^2$, respectively.

\subsection{Governing equations}

\begin{figure*}
    \begin{center}
        \includegraphics[width=\textwidth]{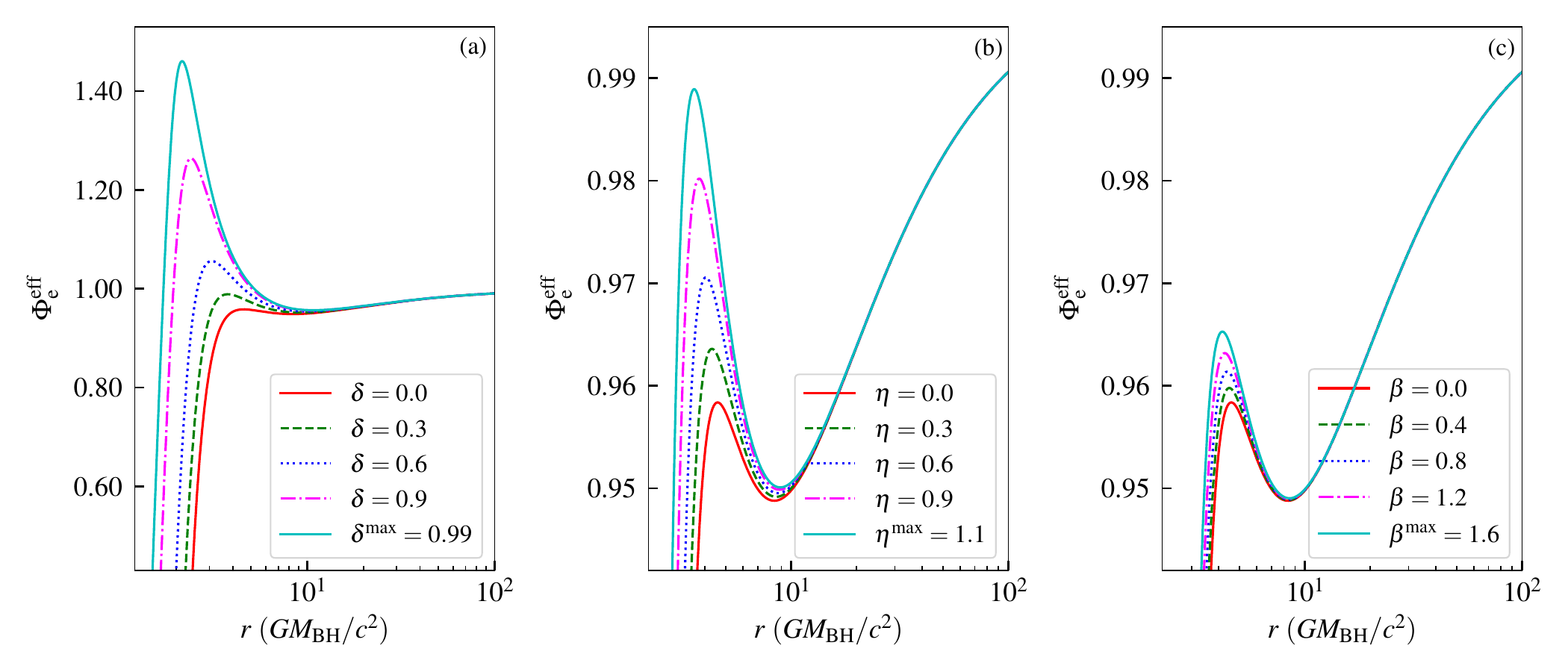}
    \end{center}    
    \caption{Variation of the effective potential $\Phi^{\rm eff}_{\rm e}$ with radial coordinate $r$. Panel (a) shows the dependence on the parameter $\delta$ with $\eta = \beta = 0$. Panels (b) and (c) illustrate the effects of the parameters $\eta$ (with $\delta = \beta = 0$) and $\beta$ (with $\delta = \eta = 0$), respectively. In each panel, different curves correspond to different values of the spacetime parameters as indicated in the legends. Here, we choose the specific angular momentum as $\lambda = 3.8$. See the text for details.
    }
    \label{fig-01}
\end{figure*}

In general relativistic hydrodynamics, the conservation of energy-momentum tensor and the mass flux yields all the hydrodynamical equations describing the flow. The energy-momentum conservation equation is given by,
\begin{equation} \label{energy_conserve_eq}
	T^{\mu\nu}_{;\nu} = 0,
\end{equation}
and the mass conservation equation is written as,
\begin{equation} \label{mass_conserve_eq}
	(\rho u^{\mu})_{;\nu} = 0.
\end{equation}
Here, $T^{\mu\nu}$ denotes the energy-momentum tensor, $\rho$ is the local mass density, and $u^{\mu}$ are the four velocities of any fluid element, respectively. The timelike velocity field satisfies the local normalization condition $u_{\mu}u^{\mu} = -1$. The energy-momentum tensor for the non-dissipative fluid composed of ions and electrons is expressed as, 
\begin{equation}
    T^{\mu\nu} = (e + P)u^{\mu}u^{\nu} + P g^{\mu\nu},
\end{equation}
where $e$, $P$, and $u^{\mu}$ denote the energy density, pressure, and four velocity vector, respectively. By projecting the conservation equation onto the spatial hypersurface using the projection operator $h^{\alpha}_{\mu} = \delta^{\alpha}_{\mu} + u^{\alpha}u_{\mu}$, we obtain the Euler equation which is given by,
\begin{align} \label{eulereqn}
    h^{\alpha}_{\mu} T^{\mu\nu}_{;\nu} 
    & = (e+P)u^{\nu}u^{\alpha}_{;\nu} + (g^{\alpha\nu} + u^{\alpha}u^{\nu})P_{,\nu} = 0.
\end{align}
It is worth noting that the condition $h^{\alpha}_{\mu} u^{\mu} = 0$ ensures the orthogonality between the projection operator and the four velocity. By projecting the conservation equation along the direction of the four velocity $u_{\mu}$, we obtain an expression representing the relativistic form of the first law of thermodynamics as,
\begin{equation} \label{firstlawthermo}
    u_{\mu} T^{\mu\nu}_{;\nu} = u^{\mu}\Bigg[\Bigg(\frac{e+P}{\rho}\Bigg)\rho_{,\mu} - e_{,\mu}\Bigg] = 0.
\end{equation}

In this analysis, we model a geometrically thin, axisymmetric accretion disk confined to the equatorial plane ($\theta = \pi/2$), where the polar component of the four velocity vanishes ($u_{\theta} = 0$). The flow dynamics are described in the co-rotating reference frame, where the three velocity is expressed as $v^{2} = \gamma_{\phi}^{2} v_{r}^{2}$. Here, the radial velocity component is defined as $v_{r}^{2} = u^{r}u_{r}/(-u^{t}u_{t})$, and the azimuthal Lorentz factor is $\gamma_{\phi}^{2} = 1/(1 - v_{\phi}^{2})$. The radial Lorentz factor is $\gamma_{v}^{2} = 1/(1 - v^{2})$, and the azimuthal velocity satisfies $v_{\phi}^{2} = u^{\phi}u_{\phi}/(-u^{t}u_{t})$. Substituting these relations into Eq. \eqref{eulereqn} and setting the index $\alpha = r$, we obtain the expression for the relativistic radial momentum equation as,
\begin{equation} \label{radialeqn}
    v \gamma^{2}_{v} \frac{dv}{dr} + \frac{1}{h\rho}\frac{dP}{dr} + \frac{d\Phi^{\rm eff}_{\rm e}}{dr} = 0,
\end{equation}
where $h ~[= (e+P)/\rho]$ is the specific enthalpy, and $\Phi^{\rm eff}_{\rm e}$ denotes the effective potential \cite[]{dihingia2018limitations} in the equatorial plane ($\theta = \pi/2$) of the disk which is given by,
\begin{equation} \label{diskpotential}
     \Phi^{\rm eff}_{\rm e} = 1 + \frac{1}{2} \ln \Bigg[\frac{-g_{tt}g_{\phi\phi}}{(g_{\phi\phi}+\lambda^{2}g_{tt})}\Bigg]
     = 1 + \frac{1}{2} \ln \Bigg[\frac{r^2\mathscr{F}(r)}{r^2 - \lambda^{2}\mathscr{F}(r)}\Bigg].\\
\end{equation}
Our consideration of static and axisymmetric spacetime implies that there exist two mutually perpendicular killing vectors that facilitate the construction of two conserved quantities along the direction of the flow, which are given by,
\begin{equation} \label{conserved_quan_kill}
    hu_{\phi} = \mathcal{L} ~~\text{(constant)}; ~~~~  -hu_{t} = \mathscr{E} ~~\text{(constant)},
\end{equation}
where $\mathscr{E}$ is the specific energy (also known as the Bernoulli constant) of the flow, and $\mathcal{L}$ corresponds to the bulk angular momentum per unit mass of the flow. In equation \eqref{conserved_quan_kill}, the index $\alpha = t,\phi$ manifests conserved energy and angular momentum, respectively. In addition, we express the specific angular momentum of the flow as $\lambda = \mathcal{L}/\mathscr{E} = -u_{\phi}/u_{t}$, which remains conserved all throughout for an inviscid flow.

With this framework, we examine how the spacetime parameters $\delta$, $\eta$ and $\beta$ influence the overall behaviour of the effective potential ($\Phi^{\rm eff}_{\rm e}$) around the central object. The obtained results are depicted in Fig. \ref{fig-01}, where the variation of $\Phi^{\rm eff}_{\rm e}$ is depicted as a function of radial coordinate $(r)$ for flows characterized by an angular momentum of $\lambda = 3.8$. In panel (a), we set $\eta = \beta = 0$, and vary the parameter $\delta$ from $0.0$ to $0.99$, in steps of $0.3$ followed by $\delta^{\rm max} = 0.99$, which is the maximum value of $\delta$ yielding BH with a horizon. We observe that $\Phi^{\rm eff}_{\rm e}$ increases for higher values of $\delta$ in the vicinity of the horizon. However, at large $r$, $\Phi^{\rm eff}_{\rm e}$ tends to merge for all values of $\delta$ with the Schwarzschild potential, indicating asymptotically flat vacuum solutions. Similarly, panels (b) and (c) illustrate the dependence of $\Phi^{\rm eff}_{\rm e}$ on $r$ for increasing values of $\eta$ (with $\delta = \beta = 0$) and $\beta$ (with $\delta = \eta = 0$), respectively. As in panel (a), we identify the results corresponding to the maximum values $\eta^{\rm max} = 1.1$ (for $\delta = \beta = 0$) and $\beta^{\rm max} = 1.6$ (for $\delta = \eta = 0$) in panels (b) and (c), respectively. A comparison of panels (a), (b), and (c) indicates that the effective potential $\Phi^{\rm eff}_{\rm e}$ remains more sensitive to variations in the parameter $\delta$ compared to changes in $\eta$ and $\beta$ parameters. Note that $\delta^{\rm max}$, $\eta^{\rm max}$, and $\beta^{\rm max}$ denote the maximum values of the spacetime parameters representing a BH, and we identify each of them by setting the other two parameters equal to zero ensuring the existence of a horizon (see Appendix \ref{eventhorizon} for details).

Thereafter, we use equation \eqref{firstlawthermo} to derive the radial component of the entropy generation equation as,
\begin{equation}
    \left(\frac{e+P}{\rho}\right)\frac{d\rho}{dr} - \frac{de}{dr} = 0.
\end{equation}
Next, we integrate equation \eqref{mass_conserve_eq}, corresponding to mass conservation, to derive the associated constant of motion, namely the mass accretion rate $(\dot{M})$ as, 
\begin{equation} \label{mass_acc}
    \dot{M} = 4 \pi r \rho v \gamma_v H\sqrt{\mathscr{F}(r)},
\end{equation}
where $H=\sqrt{P r^{3}/(\rho \gamma^{2}_\phi)}$ is the local half thickness of the disk \cite[]{riffert1995relativistic,peitz1997viscous}. 

To solve the governing equations of the flow, an appropriate relation between the pressure ($P$), mass density ($\rho$), and internal energy ($e$), namely the equation of state (EoS), is essential. Because of the intense gravitational field of the BH, the accretion flow in the vicinity of the horizon attains temperatures of about $10^{10}-10^{11}$ K, rendering the plasma inherently thermally relativistic in this region. Therefore, following \citet{chatto2009effects}, we adopt a relativistic EoS with a variable adiabatic index ($\Gamma$) to relate the thermodynamic quantities of the flow, which is expressed as,
\begin{equation} \label{eos}
    e = \frac{\rho f}{\Big(1 + \frac{m_{\rm p}}{m_{\rm e}}\Big)}, \qquad P = \frac{2\rho\Theta}{\Big(1 + \frac{m_{\rm p}}{m_{\rm e}}\Big)},
\end{equation}
with
\begin{equation*}
    f = \Bigg[ 1 + \Theta\Bigg( \frac{9 \Theta + 3}{3 \Theta +2} \Bigg) \Bigg] +
    \Bigg[ \frac{m_{\rm p}}{m_{\rm e}} + \Theta \Bigg( \frac{9 \Theta m_{\rm e} + 3 m_{\rm p}}{3 \Theta m_{\rm e} + 2 m_{\rm p}} \Bigg) \Bigg].
\end{equation*}
Here, $m_{\rm e}$ and $m_{\rm p}$ denote the electron and ion masses, respectively, and $\Theta ~(= k_{\rm B}T/(m_{\rm e} c^{2}))$ represents the dimensionless temperature, where $k_{\rm B}$ is the Boltzmann constant and $T$ is the flow temperature in Kelvin. Using the relativistic EoS, we express the adiabatic index $(\Gamma)$ and sound speed $(a_{\rm s})$ as \cite[]{chatto2009effects,dihingia2019low,uniyal2024,Sen-etal2022},
\begin{equation}\label{delineate}
    \Gamma = 1+\frac{2}{(df/d\Theta)}, \qquad a^{2}_{\rm s} = \frac{\Gamma P}{e + P} = 
    \frac{2\Gamma\Theta}{f + 2\Theta}.
\end{equation}

Thereafter, we define the entropy accretion rate of the flow as \cite[]{chatto2016estimation},
\begin{equation}\label{ent_acc}
    \dot{\mathscr{M}} = \frac{\dot{M}}{4 \pi \mathcal{K}} = \frac{\rho}{\mathcal{K}} H r v \gamma_v \sqrt{\mathscr{F}(r)},
\end{equation}
where density ($\rho$) is computed by integrating equation \eqref{firstlawthermo} using equation \eqref{eos} as,
\begin{multline}
    \rho = \mathcal{K} \exp\Bigg(\frac{f-(1 + m_{\rm p}/m_{\rm e})}{2\Theta}\Bigg) \\  \times \Theta^{3/2} (3\Theta + 2)^{3/4} (3\Theta + 2 m_{\rm p}/m_{\rm e})^{3/4},
\end{multline}
where the constant $\mathcal{K}$ represents the measure of entropy. Furthermore, for a non-dissipative flow, the entropy accretion rate $\dot{\mathscr{M}}$ remains constant along the flow streamline \cite[][and references therein]{dihingia2020study}.

\subsection{Wind Equation}

We use equations \eqref{radialeqn}, \eqref{mass_acc} and \eqref{eos}, to obtain the radial derivative of the flow velocity in the form of the wind equation as,
\begin{equation} \label{dvdr}
    \frac{dv}{dr} = \frac{\mathcal{N}}{\mathcal{D}},
\end{equation}
where the numerator $\mathcal{N}$ is given by,
\begin{equation} \label{num}
    \mathcal{N} = \frac{2 a^{2}_{s}}{\Gamma + 1} \Bigg[ \frac{5}{2r} + \frac{1}{2 \mathscr{F}(r)} 
    \frac{d\mathscr{F}(r)}{dr} 
    - \frac{1}{2\gamma^{2}_{\phi}} \frac{d\gamma^{2}_{\phi}}{dr} \Bigg] - \frac{d\Phi^{\rm eff}_{\rm e}}{dr},
\end{equation}
and the denominator is given by,
\begin{equation} \label{den}
    \mathcal{D} = \gamma^{2}_{v} \Bigg( v - \frac{2 a^{2}_{\rm s}}{v(\Gamma + 1)} \Bigg)
\end{equation}
Further, we obtain the gradient of the dimensionless temperature using equations \eqref{eulereqn}, \eqref{mass_acc} and  \eqref{delineate} as,
\begin{equation} \label{dtdr}
    \frac{d\Theta}{dr} = -\frac{2\Theta}{2N + 1} \Bigg[ \frac{\gamma^{2}_{v}}{v} \frac{dv}{dr}
    + \frac{5}{2r} + \frac{1}{2 \mathscr{F}(r)} 
    \frac{d\mathscr{F}(r)}{dr} - \frac{1}{2\gamma^{2}_{\phi}} \frac{d\gamma^{2}_{\phi}}{dr} \Bigg].
\end{equation}

\subsection{Critical Point Analysis}

\begin{figure*}
    \begin{center}
        \includegraphics[width=\textwidth]{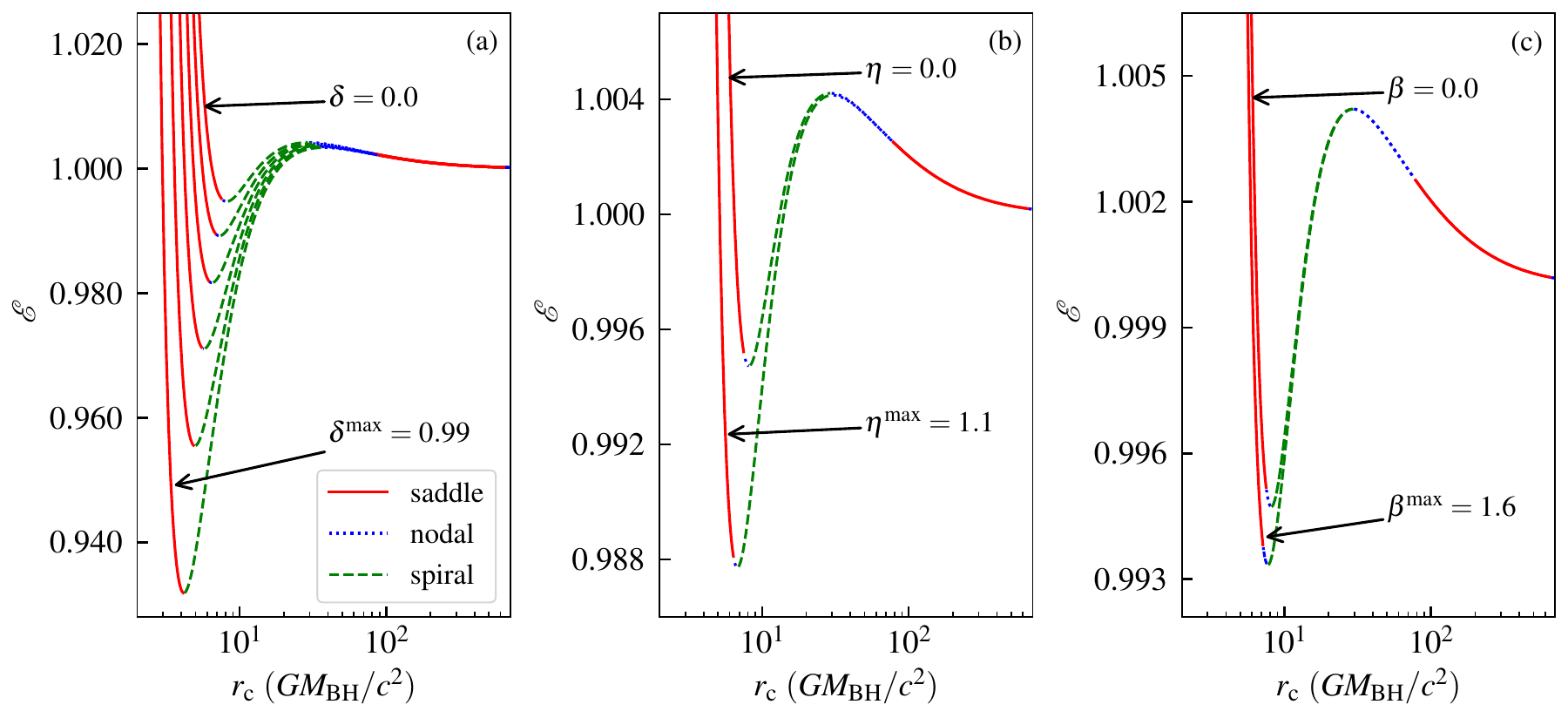}
    \end{center}    
    \caption{Variation of specific energy $\mathscr{E}$ as a function of the critical point location $r_{\rm c}$ for different choices of the spacetime parameters $\delta$, $\eta$, and $\beta$. Here, the specific angular momentum is fixed at $\lambda = 3.0$. Panel (a) illustrates the effect of varying $\delta$ ($\Delta \delta = 0.2$) with $\eta = \beta = 0$, while panels (b) and (c) display the variations resulting from changing $\eta$ (with $\delta = \beta = 0$) and $\beta$ (with $\delta = \eta = 0$), respectively. In each subplot, the solid (red), dotted (blue), and dashed (green) segments of the curves correspond to saddle, nodal, and spiral critical points. See the text for details.
    }
    \label{fig-02}
\end{figure*}

In the process of accretion onto a BH, the inflowing matter begins its journey from the outer edge of the disk ($r_{\rm edge}$) with an initially subsonic radial velocity ($v \sim 0$). As the flow moves inward, the intense gravitational field in the vicinity of the BH horizon causes its radial velocity to increase steadily, eventually driving the flow through a sonic state transition. This transition occurs smoothly at the critical point ($r_{\rm c}$), where the flow becomes supersonic in order to satisfy the inner boundary condition imposed at the horizon. At $r_{\rm c}$, both numerator and denominator of the radial velocity gradient (see equation \eqref{dvdr}) simultaneously vanish ($i.e.$, $\mathcal{N} = 0 = \mathcal{D}$), resulting in the indeterminate form $dv/dr \rightarrow 0/0$.

The accretion flow must remain smooth, which requires the radial velocity gradient $dv/dr$ to remain real and finite along the streamline. To satisfy this regularity condition, we evaluate the gradient $(dv/dr)_{r_{\rm c}}$ at the critical point ($r_{\rm c}$), by applying l$'$H\^{o}pital's rule. This yields two possible values of $(dv/dr)_{r_{\rm c}}$, and based on these values, critical points are classified into three types: saddle, nodal, and spiral. A saddle type critical point occurs when the two values of $(dv/dr)_{r_{\rm c}}$ are real and of opposite sign; a nodal critical point arises when both values are real and share the same sign; and in the case of a spiral critical point, the two values become complex \cite[][and references therein]{liang1980transonic, chakrabarti1989standing, das2001standing, das2007behaviour, kumar2013effect}. Among these categories, only the saddle type critical point permits a physically acceptable transonic accretion solution \cite[]{kato1993sonic}, where the negative value of $(dv/dr)_{r_{\rm c}}$ corresponds to the accretion branch, and the positive value represents the wind branch.

By applying the critical point conditions ($\mathcal{N} = 0 = \mathcal{D}$), we solve equation \eqref{conserved_quan_kill} for a set of model parameters. To examine the transonic behaviour of the accretion flow, we compute the specific energy ($\mathscr{E}$) as a function of the critical point location ($r_{\rm c}$) for a chosen set of $\delta$, $\eta$, and $\beta$ values, considering the flow angular momentum $\lambda = 3.0$. The obtained results are presented in Fig. \ref{fig-02}, which illustrates the variation of $\mathscr{E}$ with $r_{\rm c}$. In Fig. \ref{fig-02}a, we show the results for different $\delta$ values, starting from $\delta = 0$ in steps of $\Delta \delta = 0.2$, while keeping $\eta = \beta = 0$. For each choice of $\delta$, the critical points appear in the sequence of \textit{saddle–spiral–nodal–saddle} as $r_{\rm c}$ increases. Furthermore, we find that a range of $\mathscr{E}$ and $\delta$ exists for which the accretion flow admits multiple saddle-type critical points, which is essential for the formation of shocks waves \citep{chakrabarti1989standing, das2001standing, Das2004shock, das2007behaviour, dihingia2019shocks, mitra2024mhdshock}. Needless to mention that $\delta = 0$ case in Fig. \ref{fig-02}a corresponds to the Schwarzschild spacetime. Similarly, Figs. \ref{fig-02}b and \ref{fig-02}c show the results for varied $\eta$ and $\beta$, where we adopt $\delta = \beta = 0$ and $\delta = \eta = 0$, respectively. We observe that the overall influence of $\eta$ and $\beta$ on the dependence of $\mathscr{E}$ on $r_{\rm c}$ is comparatively weaker than that of the parameter $\delta$.

\begin{figure*}
    \begin{center}
        \includegraphics[width=\textwidth]{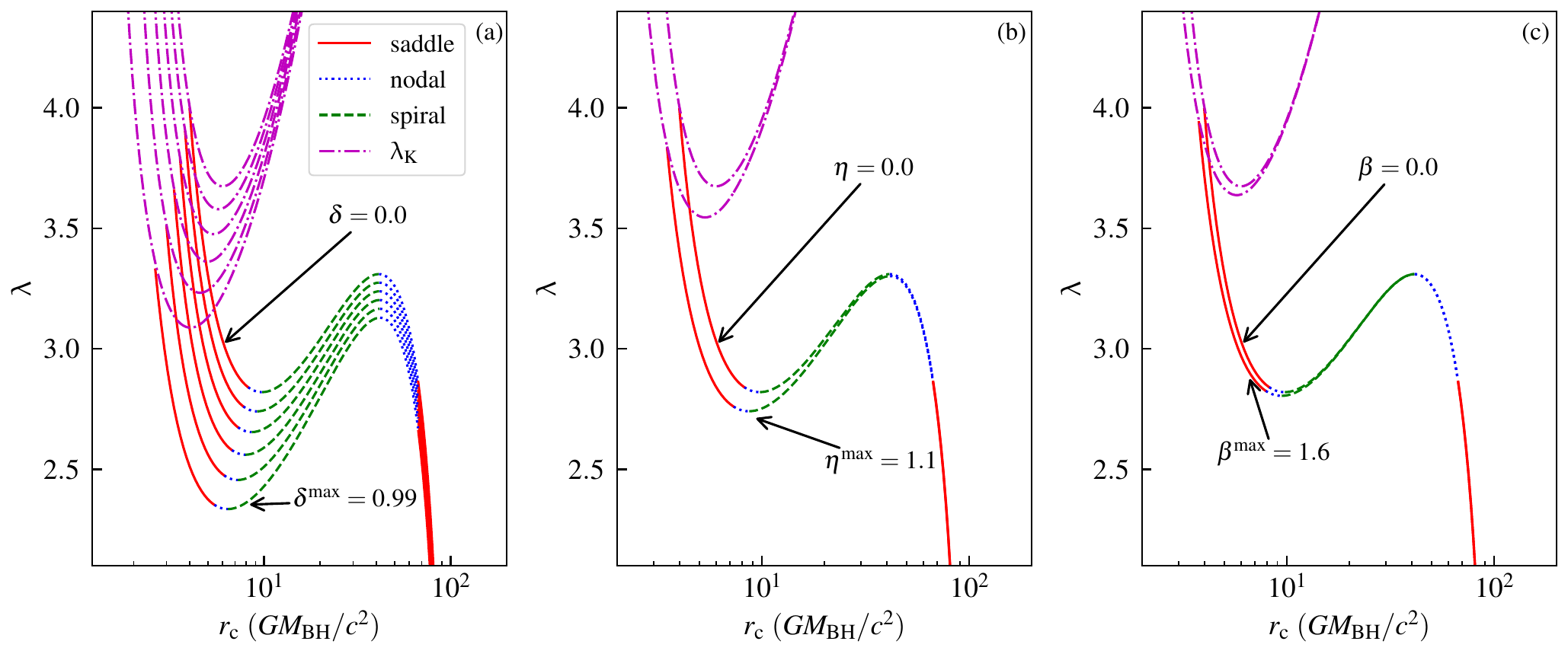}
    \end{center}
    \caption{Variation of specific angular momentum $\lambda$ as a function of the critical point location $r_{\rm c}$ for different values of the spacetime parameters $\delta$, $\eta$, and $\beta$. Here, we keep the specific energy fixed at $\mathscr{E} = 1.003$. Panel (a) presents the influence of varying $\delta$ (with $\Delta \delta =0.2$) while keeping $\eta = \beta = 0$, whereas panels (b) and (c) show the corresponding variations resulting from changes in $\eta$ (with $\delta = \beta = 0$) and in $\beta$ (with $\delta = \eta = 0$), respectively. In all panels, solid (red), dotted (blue), and dashed (green) segments of each curve denote saddle, nodal, and spiral critical points, respectively. The dot–dashed (magenta) curves represent the Keplerian angular momentum distribution $\lambda_{\rm K}$. See the text for details.
    }
    \label{fig-03}
\end{figure*}

We further examine the transonic properties of the accretion flow in Fig. \ref{fig-03}, where the variation of the specific angular momentum ($\lambda$) is plotted as function of the critical point ($r_{\rm c}$) for different choices of the spacetime parameters. In panel (a), $\lambda(r_{\rm c})$ is shown for a sequence of $\delta$ values with an interval of $\Delta \delta = 0.2$, while panels (b) and (c) display the corresponding variations for different values of $\eta$ and $\beta$, respectively. Throughout the figure, we fix flow energy as $\mathscr{E} = 1.003$. Following the convention used in Fig. \ref{fig-02}, the solid (red), dotted (blue), and dashed (green) segments of each curve denote saddle, nodal, and spiral critical points, respectively. To facilitate comparison with the Keplerian distribution, we compute the Keplerian angular momentum ($\lambda_{\rm K}$) by imposing the condition $d\Phi^{\rm eff}_{\rm e}/dr = 0$ at each $r_{\rm c}$. The resulting $\lambda_{\rm K}(r_{\rm c})$ profile is over plotted in all panels using the dot–dashed (magenta) curve. The comparison clearly indicates that the flow remains sub-Keplerian over the entire radial extent between the event horizon ($r_{\rm H}$) and the outer edge of the disk ($r_{\rm edge}$), irrespective of the choice of model parameters. In addition, panel (a) demonstrates that increasing $\delta$ value exerts the intense influence on the angular momentum distribution and the nature of the critical points exhibits noticeable deviation from the Schwarzschild limit ($\delta = 0$). The results in panels (b) and (c) evidently show that the effects of $\eta$ and $\beta$ are comparatively weaker, consistent with the trends observed in the earlier analysis of $\mathscr{E}(r_{\rm c})$. Overall, the obtained results confirm that $\delta$ plays the dominant role in deciding the transonic properties of the accretion flow. This is expected as higher-order corrections to the metric components compared to the $1/r^2$ dependency results in subdominant impact on the accretion flow properties. In a way, faster decaying higher-order terms seem to contribute at the immediate vicinity of the compact object and therefore provide localized corrections that do not substantially alter the transonic properties of the flow.

\begin{figure}
    \begin{center}
         \includegraphics[width=\columnwidth]{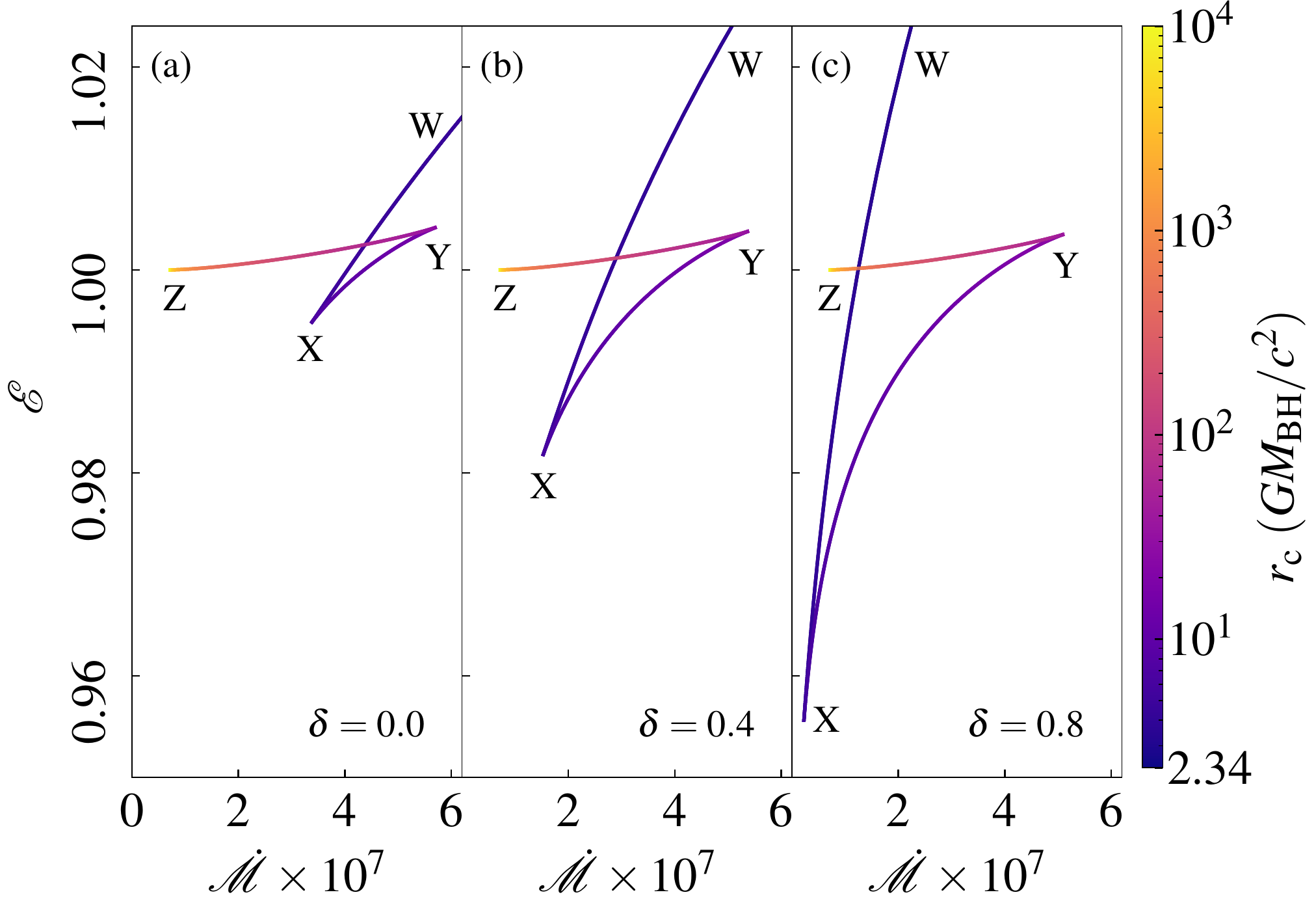}
    \end{center}
    \caption{Variation of specific energy $\mathscr{E}$ with the entropy accretion rate $\dot{\mathscr{M}}$ for flow with fixed specific angular momentum $\lambda = 3.0$. In panels (a), (b), and (c), the parameter $\delta$ is set to $0.0$, $0.4$, and $0.8$, respectively, while keeping $\eta = \beta = 0$. The colour gradient represents the critical point location $r_{\rm c}$ associated with variations in $\mathscr{E}$ and $\dot{\mathscr{M}}$. The plot distinctly shows that all points along the WX branch correspond to inner critical points, the YZ branch corresponds to outer critical points, and the XY branch provides the spiral critical points. See the text for details.
    }
    \label{fig-04}
\end{figure}

To complement the critical point analysis, we explore the roles the specific energy $\mathscr{E}$ and entropy accretion rate $\dot{\mathscr{M}}$ that render critical points. In doing so, we calculate $\mathscr{E}$ (equation \eqref{conserved_quan_kill}) and $\dot{\mathscr{M}}$ (equation \eqref{ent_acc}) at the critical point for $r_{\rm H} < r_{\rm c} \le r_{\rm edge}$. Here, we choose $\lambda = 3.0$ and set $\eta = \beta =0$. We depict the obtained results in Fig. \ref{fig-04} for different values of $\delta$, which is marked in each panel. In panel (a) of the figure, the parameters from WX branch result in accretion solutions passing through inner critical point ($r_{\rm in}$), whereas the parameters from YZ branch yield solutions passing through outer critical points ($r_{\rm out}$). It is worth noting that parameters on XY branch provide spiral critical points which are unphysical ($(dv/dr)_{r_{\rm c}}$ becomes complex) as flow can not pass through these points. At the intersection of WX and YZ branches, the entropy accretion rates at $r_{\rm in}$ and $r_{\rm out}$ become equal, and we denote this critical value and its corresponding energy as $\dot{\mathscr{M}}_{m}$ and $\mathscr{E}_{m}$, respectively. For the energy range $\mathscr{E}_{Y} \le \mathscr{E} \le \mathscr{E}_{Z}$, the flow admits multiple saddle-type critical points ($r_{\rm in}$ and $r_{\rm out}$). Across this interval, the entropy accretion rates at the two critical points depends on $\mathscr{E}$. In particular, for $\mathscr{E}_{Z} \le \mathscr{E} \le \mathscr{E}_{m}$, the entropy accretion rate $\dot{\mathscr{M}}_{\rm in} \ge \dot{\mathscr{M}}_{\rm out}$, whereas the opposite inequality ($\dot{\mathscr{M}}_{\rm in} \le \dot{\mathscr{M}}_{\rm out}$) holds for $\mathscr{E}_{m} \le \mathscr{E} \le \mathscr{E}_{Y}$. Here, the subscripts ‘in’ and ‘out’ refer to quantities measured at $r_{\rm in}$ and $r_{\rm out}$, respectively. It is noteworthy that non-zero $\delta$ values yield qualitatively similar results (see panel (b) and (c)), however, the overall range of WX and XY increases, which eventually renders the multiple critical points for wider range of energy ($\mathscr{E}$) and entropy accretion rate ($\dot{\mathscr{M}}$).

\begin{figure}
    \begin{center}
        \includegraphics[width=\columnwidth]{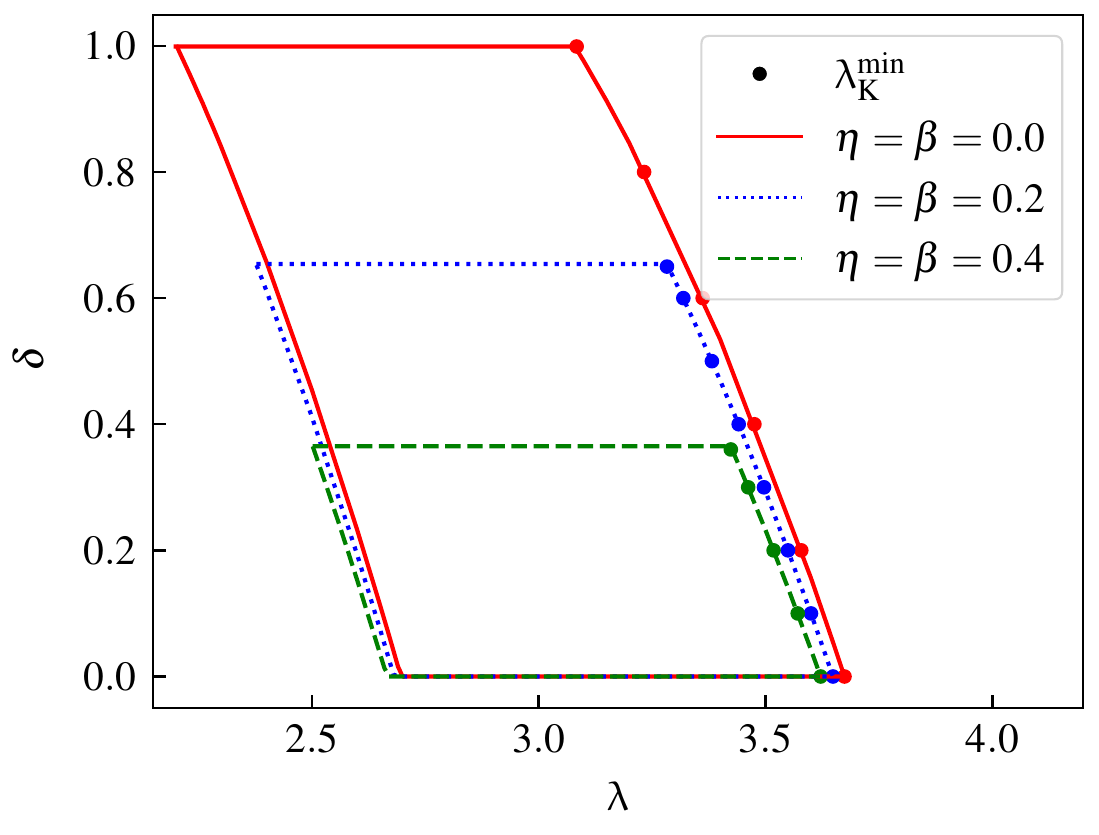}
    \end{center}
    \caption{Region of $\delta–\lambda$ parameter space for different combinations of the spacetime parameters $\eta$ and $\beta$ that admit multiple critical points. The regions enclosed by the solid (red), dotted (blue) and dashed (green) curves are obtained for the parameter sets $(\eta,\beta) = (0,0)$, $(0.2,0.2)$, and $(0.4,0.4)$, respectively. The filled circles on each curve mark the corresponding minimum Keplerian angular momentum values $\lambda^{\rm min}_{\rm K}$ for a given choice of $(\delta,\eta,\beta)$. See the text for details.
    }
    \label{fig-05}
\end{figure}

Next, we examine the influence of the spacetime parameters $\delta$, $\eta$, and $\beta$ in determining the accessible range of angular momentum ($\lambda$) that permits the formation of multiple critical points. The resulting parameter space spanned in the $\delta-\lambda$ plane is illustrated in Fig. \ref{fig-05}, where three representative combinations $(\eta,\beta) = (0.0,0.0)$, $(0.2,0.2)$, and $(0.4,0.4)$ are shown. Here, we do not fix the specific energy $\mathscr{E}$ of the flow; instead, we vary it freely. For each case, the left boundary of the enclosed region marks the minimum angular momentum allowed for the corresponding set of $(\delta,\eta,\beta)$, whereas the lower boundary identifies the range of $\lambda$ that yields multiple critical points at $\delta = 0$ for the chosen values of $\eta$ and $\beta$. Although the individual effects of $\eta$ and $\beta$ on the critical points are comparatively weaker than that of $\delta$, these parameters nonetheless constrain the maximum value of $\delta$ permissible for BH accretion. These findings are reflected in the upper boundary of each region, which represents the corresponding $\delta^{\rm max}$. The right boundary denotes the limiting angular momentum for a given set of $(\delta,\eta,\beta)$ for which multiple critical points exist while the flow remains sub-Keplerian. The filled points along this edge trace the minimum angular momentum ($\lambda^{\rm min}_{\rm K}$) of a Keplerian distribution for the given $(\delta,\eta,\beta)$ configuration (see Appendix \ref{Keplerian}).

\section{Accretion Solutions in Generic Spacetime}
\label{sec:accretion}

In this section, we investigate the nature of transonic global accretion solutions in the spacetime described by the generic metric adopted in this study. Our analysis focuses on how the underlying spacetime geometry affects the overall transonic behaviour of the flow. In particular, we examine the role of the parameter $\delta$, which encapsulates deviations from the standard BH spacetime, and assess how it modifies the range of flow parameters that yield physically consistent transonic accretion solutions. By systematically varying the parameter $\delta$, we aim to identify the different types of global transonic accretion solutions.

\begin{figure*}
    \begin{center}
        \includegraphics[width=1.0\textwidth]{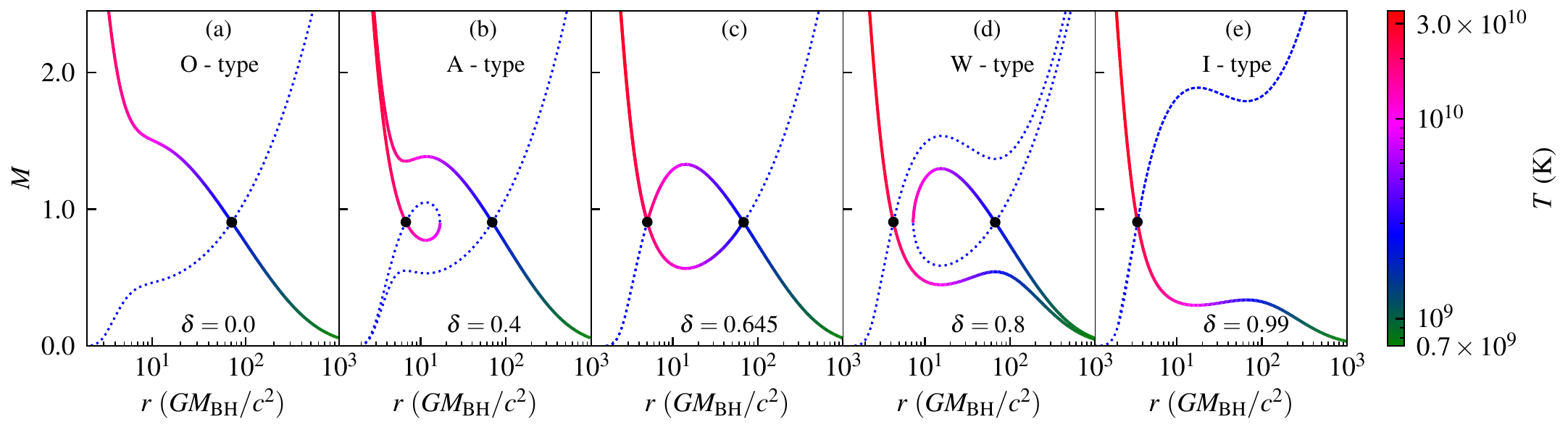}
    \end{center}
    \caption{Alteration of transonic flow solutions with increasing spacetime parameter $\delta$. Each panel shows the variation of Mach number $M$ as a function of radial coordinate $r$ for flows with $\mathscr{E} = 1.003$ and $\lambda = 2.7$, with $\eta = \beta = 0$. Dotted (blue) curves depict the wind branch, and the colour along the accretion branch indicates the corresponding temperature variation. The sequence from left to right demonstrates the progressive change in solution topology from panel (a) O-type to panel (e) I-type as $\delta$ is increased. Panels (b) and (d) display the intermediate A-type (outer branch open, inner closed) and W-type (inner open, outer closed) solutions, while panel (c) denotes the transitional case where both critical points admit identical entropy accretion rates ($\dot{\mathscr{M}}_{\rm in} = \dot{\mathscr{M}}_{\rm out}$). See the text for details.
    }
    \label{fig-06}
\end{figure*}

\subsection{Global Transonic Solutions}

In order to obtain the transonic accretion solutions, we simultaneously solve equations \eqref{dvdr} and \eqref{dtdr} for a chosen set of input parameters, namely $\mathscr{E}$, $\lambda$, $\delta$, $\eta$, and $\beta$. As a first step, we perform the critical point analysis by determining the critical radius $r_{\rm c}$ and the corresponding velocity gradient $(dv/dr)_{r_{\rm c}}$ through the solution of equation \eqref{conserved_quan_kill}. Once the critical point is identified, we integrate equations \eqref{dvdr} and \eqref{dtdr} both inward up to the event horizon ($r_{\rm H}$), and then outward toward the outer disc edge ($r_{\rm edge}$). These two parts of the solution are subsequently joined to obtain a continuous global transonic accretion flow that extends from the disc outer edge ($r_{\rm edge}$) down to the BH horizon ($r_{\rm H}$). Depending on the chosen parameter set, the resulting solutions are classified into four distinct classes, namely O-type, A-type, W-type, and I-type, each representing a characteristic topology of the transonic flow \cite[]{dihingia2019shocks}.

In Fig. \ref{fig-06}, we present the transonic global accretion solutions obtained by varying the parameter $\delta$, while keeping $\eta = \beta = 0$. This choice is motivated by the fact that the solutions are generally sensitive to variations in $\delta$, whereas the changes due to $\eta$ and $\beta$ are relatively feeble (see Figs. \ref{fig-02} and \ref{fig-03}). In the figure, the Mach number ($M=v/a_{\rm s}$) of the flow is plotted as a function of radial distance ($r$). For the solutions presented in the figure, we have considered $\mathscr{E} = 1.003$ and $\lambda = 2.7$, and vary the value of $\delta$ from left to right across the panels which are marked. In each panel, solid curve is used to denote the accretion solution, while the dotted curve refers winds, and filled circle denotes the critical point.

In panel (a), we present the O-type transonic solution that passes through the outer critical point at $r_{\rm out} = 70.8984$ for $\delta = 0$. When the value of $\delta$ is increased to $0.4$ (panel (b)), an inner critical point emerges at $r_{\rm in} = 6.6934$ in addition to the outer critical point at $r_{\rm out} = 69.0794$. We note that the accretion solution passing through $r_{\rm in}$ forms a closed branch that does not extend from the disc edge to the event horizon. In contrast, the solution passing through the outer critical point at $r_{\rm out} = 69.0794$ yields a global transonic accretion flow that smoothly connects the horizon ($r_{\rm H}$) to the outer edge of the disk ($r_{\rm edge}$). For this case, we obtain $\dot{\mathscr{M}}_{\rm in} = 5.90 \times 10^7$ and $\dot{\mathscr{M}}_{\rm out} = 4.97\times 10^7$, corresponding to the well-known A-type class of solutions \cite[]{dihingia2019shocks}. At a particular value of $\delta = 0.645$ (panel (c)), both $r_{\rm in}$ and $r_{\rm out}$ admit solutions with equal entropy accretion rates, $\dot{\mathscr{M}}_{\rm in} = \dot{\mathscr{M}}_{\rm out} = 4.94 \times 10^7$. For $\delta = 0.8$ (panel (d)), the nature of the solutions reverses, $i.e.$ the branch passing through $r_{\rm in} = 4.2825$ now becomes the global solution that connects $r_{\rm H}$ to $r_{\rm edge}$, whereas the branch possessing the outer critical point at $r_{\rm out} = 67.1213$ becomes closed. Here, we obtain $\dot{\mathscr{M}}_{\rm in} = 4.14 \times 10^7$ and $\dot{\mathscr{M}}_{\rm out} = 4.93 \times 10^7$, which is the characteristic of W-type solutions \citep{dihingia2019shocks}. When $\delta$ is increased further to $0.99$ (panel (e)), the outer critical point disappears entirely, leaving only the inner critical point at $r_{\rm in} = 3.4498$ (I-type solution), for which $\dot{\mathscr{M}}_{\rm in} = 2.89 \times 10^7$.

\subsection{Parameter Space for Multiple Critical Points}

\begin{figure}
    \begin{center}
        \includegraphics[width=\columnwidth]{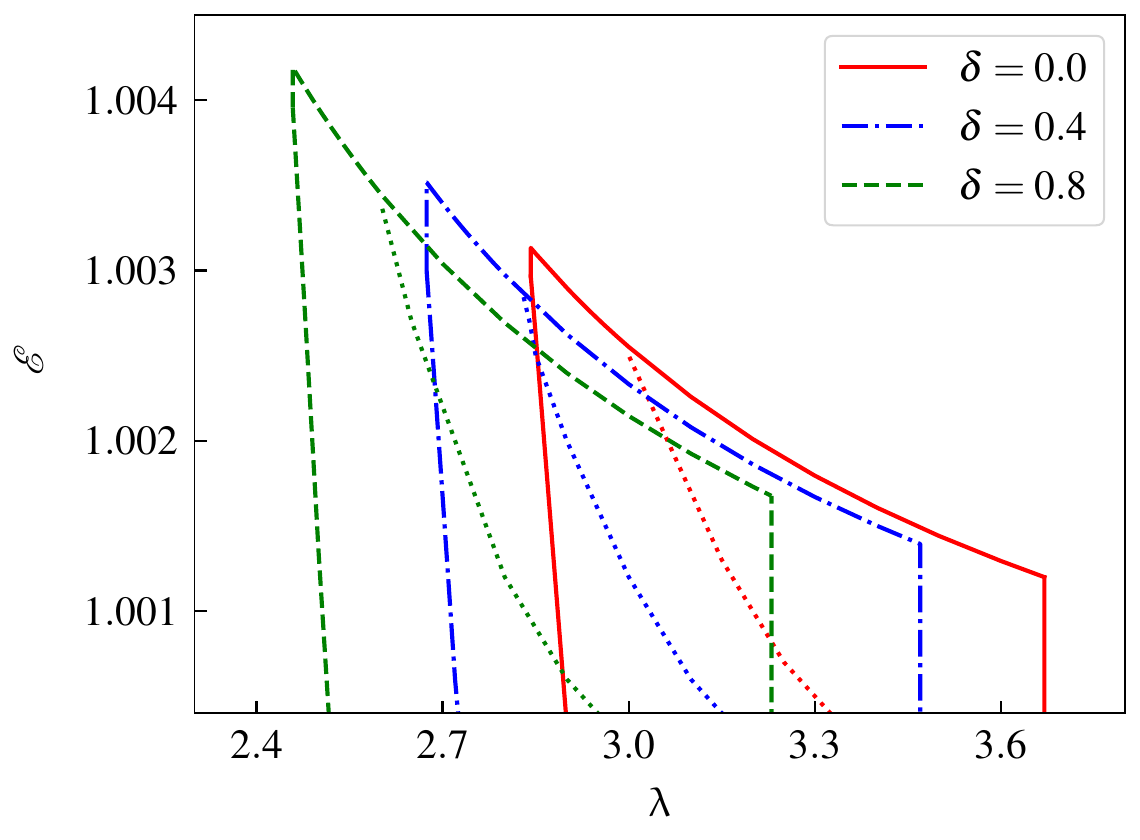}
    \end{center}
    \caption{Modification of the $\lambda-\mathscr{E}$ parameter space for increasing $\delta$ for fixed $\eta = \beta = 0$. The region bounded by the solid (red) curve corresponds to $\delta = 0$ (the Schwarzschild BH limit), while the dot–dashed (blue) and dashed (green) curves represent $\delta = 0.4$ and $0.8$, respectively. For each $\delta$, the dotted curve ($\dot{\mathscr{M}}_{\rm in} = \dot{\mathscr{M}}_{\rm out}$) divides the corresponding region into domains that admit A-type and W-type accretion solutions. See the text for details.
    }
    \label{fig-07}
\end{figure}

Since the angular momentum $\lambda$ and specific energy $\mathscr{E}$ of the accretion flow determine the nature and number of critical points, it is essential to delineate the region of parameter space that admits multiple critical points.  In Fig. \ref{fig-07}, we present the $\lambda-\mathscr{E}$ parameter space for different values of the spacetime parameter $\delta$, setting $\eta = \beta = 0$. In the figure, the effective domain of the parameter space separated by the red, blue and green curves are obtained for $\delta = 0$, $0.4$ and $0.8$, respectively. Figure evidently shows that increasing $\delta$ systematically shifts the admissible range of $\lambda$ toward lower angular momentum side while broadening the allowable range of $\mathscr{E}$, indicating that larger $\delta$ supports global transonic solutions at higher energies but lower angular momentum. For each $\delta$, the parameter space bifurcates into two regions separated by the dotted curve defined by $\dot{\mathscr{M}}_{\rm in}=\dot{\mathscr{M}}_{\rm out}$ (see Fig. \ref{fig-06}c). To the left of this dotted curve lie the A-type solutions with $\dot{\mathscr{M}}_{\rm in} > \dot{\mathscr{M}}_{\rm out}$, where the global transonic branch passes through the outer critical point $r_{\rm out}$. Conversely, the right hand region of the dotted curve corresponds to W-type solutions with $\dot{\mathscr{M}}_{\rm in} < \dot{\mathscr{M}}_{\rm out}$, for which the physically relevant global solution is the branch passing through the inner critical point $r_{\rm in}$ and extending smoothly from the disc edge to the event horizon.

\subsection{Emission from the Accretion Disc}

In the innermost regions of the rotating accretion flow, the strong gravitational field of the BH compresses the gas and drives it to extremely high temperatures and densities. Under such conditions, the flow is primarily composed of hot ions and electrons. The radiative emission from this region is generally dominated by thermal bremsstrahlung (free-free) processes \citep[]{quenby2010foundations, okuda2019possible}. For sufficiently hot accretion flows, where the plasma temperature can reach as high as $T \approx 10^{11} \,\mathrm{K}$ \citep[]{yuan2014hot}, the electrons become relativistic ($k_{\rm B} T_{\rm e} > m_{\rm e} c^{2}$). In this relativistic regime, electron–electron bremsstrahlung contributes significantly to the overall emissivity. Following the prescription of \cite{novikov1973}, we take into account the relativistic effect by considering the electron-electron emission along with the standard electron-ion emission while evaluating the total bremsstrahlung emissivity. An approximate expression for the emissivity at a frequency $\nu$ is therefore written as, 
\begin{multline}
    \epsilon(\nu) = \frac{32\pi e^{6}}{3 m_{\rm e} c^{3}} \Bigg( \frac{2\pi}{3 k_{\rm B} m_{\rm e}}\Bigg)^{1/2}
    n_{\rm e}^{2}~ T_{\rm e}^{-1/2} \\ \times g_{\rm br}~(1 + 4.4 \times 10^{-10}~T_{\rm e})~\exp\Bigg(\frac{-{h}\nu}{k_{\rm B} T_{\rm e}}\Bigg),
\end{multline}
where $e$, $m_{\rm e}$, and $T_{\rm e}$ are the charge, mass, and temperature of the electron, respectively. Also, $h$ is the Planck constant, $k_{\rm B}$ is the Boltzmann constant, and $g_{\rm br}$ is the Gaunt factor \cite[]{karzas1961electron} which we choose to be unity for simplicity. We estimate the
electron temperature following \cite{chattopadhyay2002radiatively} as $T_{\rm e} = \sqrt{m_{\rm e}/m_{\rm p}}~T$, where $T$ is the flow temperature. With this, we calculate the disk luminosity from free-free emission as \cite[][and references therein]{Sen-etal2022,Sen-etal2024},
\begin{equation}
    L_{\nu_{\rm o}} = 2 \int_{0}^{\infty} \int_{r_{\rm H}}^{r_{\rm edge}} \int_{0}^{2\pi} H r \epsilon(\nu_{\rm e}) d\nu_{\rm o} dr d\phi.
\end{equation}
Here, $\nu_{\rm e}$ and $\nu_{\rm o}$ imply the emitted and observed frequencies, respectively, which are related to each other by a red-shift factor $z$ as $\nu_{\rm e} = (1 + z) \nu_{\rm o}$. The red-shift factor $z$ is defined as,
\begin{equation}
    1 + z = u^t (1 + r\Omega \sin{\phi} \sin{i}),
\end{equation}
where $\Omega = u^{\phi}/u^{t}$ is the angular velocity of the flow, while $i$ is the source inclination angle. In this work, we choose a fixed inclination angle as $i = \pi/4$, and the mass of the BH as $M_{\rm BH} = 10 \,{\rm M}_\odot$, where ${\rm M}_\odot$ denotes the solar mass. In addition, we choose the mass accretion rate to be $\dot{M} = 0.1 \dot{M}_{\rm Edd}$, where $\dot{M}_{\rm Edd}~[= 1.44 \times 10^{17} \,({\rm M_{\rm BH}/M}_\odot)~{\rm g~s}^{-1}]$ is the Eddington accretion rate.

\begin{figure}
    \begin{center}
        \includegraphics[width=\columnwidth]{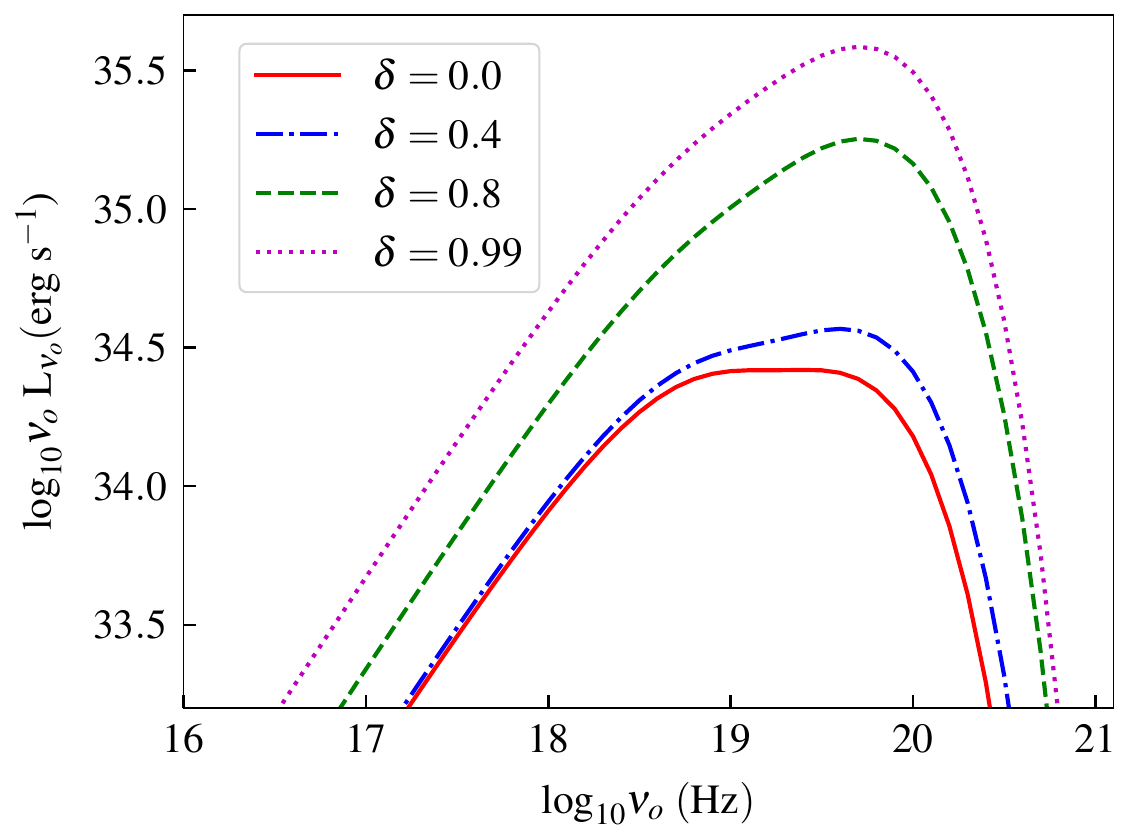}
    \end{center}    
    \caption{Variation of the spectral energy distribution (SED) with the spacetime parameter $\delta$. The solid (red) curve represents the SED for the Schwarzschild limit ($\delta = 0$), while the dot–dashed (blue), dashed (green), and dotted (magenta) curves are for $\delta = 0.4$, $0.8$, and $0.99$, respectively. These SEDs correspond to four distinct global accretion solutions shown in Figs. \ref{fig-06}(abde), which are computed for $\mathscr{E} = 1.003$, $\lambda = 2.7$, and a BH mass of $M_{\rm BH} = 10\,{\rm M}_{\odot}$. See the text for details. 
    }
    \label{fig-08}
\end{figure}

In order to connect the observable signatures with the spacetime geometry, we attempt to calculate the spectral energy distributions (SEDs) by varying the parameter $\delta$. Accordingly, in Fig. \ref{fig-08}, we present the SEDs corresponding to accretion solutions shown in Fig. \ref{fig-06}, where $\delta$ is varied while the remaining spacetime parameters are fixed at $\eta = \beta = 0$. Here, we choose $\mathscr{E} = 1.003$ and $\lambda = 2.7$. In the figure, solid (red) curve denotes the SED of O-type solution that passes through $r_{\rm out}$ (see Fig. \ref{fig-06}a) for $\delta = 0$, which eventually represents the Schwarzschild BH case. The dot-dashed (blue) curve shows the SED for $\delta = 0.4$, associated with the A-type solution in Fig. \ref{fig-06}b, where the solution passing through $r_{\rm out}$ provides the global accretion flow connecting the horizon ($r_{\rm H}$) to the disc outer edge ($r_{\rm edge}$). The dashed (green) curve illustrates the SED for $\delta = 0.8$ that corresponds to the W-type solution (Fig. \ref{fig-06}d), where the global transonic accretion flow possesses $r_{\rm in}$ and extends smoothly from $r_{\rm edge}$ to $r_{\rm H}$. Finally, the dotted (magenta) curve shows the SED for I-type solution with $\delta = 0.99$ corresponding to the accretion solution presented in Fig. \ref{fig-06}e. We find that the SEDs systematically increase with increasing values of $\delta$. This trend arises because enhanced $\delta$ leads to a reduction in the horizon radius allowing the accreting matter to approach closer to the central object. Consequently, the flow attains higher plasma temperatures near the inner edge of the disc, as evident from Fig. \ref{fig-06}. We further observe that the disc luminosity arising from bremsstrahlung emission coarsely peaks at around $\nu_0 \approx 10^{20}$ Hz, with a sharp cut-off depending on $\delta$. In reality, this high frequency cut-off directly corresponds to the electron temperature at the inner edge of the disc, which increases for higher $\delta$ as the flow reaches deeper into the gravitational potential in the vicinity of the horizon \cite[][and references therein]{Patra-etal2024}.

\section{Conclusions}\label{sec:summary}

In this work, we investigate the dynamical properties of low angular momentum accretion flows in a general class of static, spherically symmetric spacetimes characterized by the parameters $\delta$, $\eta$, and $\beta$. By solving the governing equations for a steady, axisymmetric, inviscid and advective flow, we examine how these spacetime parameters influence the structure of the effective potential, the critical point distribution, and the global transonic behaviour of the flow. With this, we summarize the key findings of this study below:

\begin{itemize}
    \item The present analysis reveals that the accretion dynamics are predominantly sensitive to the parameter $\delta$, whereas the parameters $\eta$ and $\beta$ introduce comparatively minor modifications. This behavior is consistently reflected by the trends displayed in Fig. \ref{fig-01}, Fig. \ref{fig-02}, and Fig. \ref{fig-03}. 
    
    \item We find that the parameter $\delta$ plays a decisive role in governing both the existence and the nature of critical points in low angular momentum accretion flows. As illustrated in Fig. \ref{fig-04}, the admissible range of the flow energy $\mathscr{E}$ broadens with increasing $\delta$, which enhances the possibility that flow admits multiple critical points. Subsequently, we delineate the region of the $(\delta-\lambda)$ parameter space that allows multiple critical points for flows with sub-Keplerian angular momentum ($\lambda < \lambda_{\rm K}$).

    \item We compute the complete set of transonic solutions, namely O-type, A-type, W-type and I-type, in generic class of spacetimes by systematically varying the model parameters ($\mathscr{E}$, $\lambda$, and $\delta$), for the first time to the best of our knowledge (see Fig. \ref{fig-06}). Our analysis shows that these different solution classes are not isolated solutions, instead each solution type persists over an extended region of parameter space. Considering this, we identify the effective domain in the $(\lambda-\mathscr{E})$ plane for different values of $\delta$ for solutions containing multiple critical points. These parameter spaces are further sub-divided based on the entropy accretion rate at the inner and outer critical points ($\mathscr{\dot M}_{\rm in} \lessgtr \mathscr{\dot M}_{\rm out}$). We find that the allowed $(\lambda-\mathscr{E})$ region systematically shifts toward higher energies and lower angular momenta as $\delta$ increases (see Fig. \ref{fig-07}).
    
    \item Next, we compute the spectral energy distributions (SEDs) associated with the full set of global transonic accretion solutions (O, A, W, and I-types) and find that the radiative output is regulated by the spacetime parameter $\delta$. In particular, the SEDs corresponding to the I-type solutions exhibit noticeably higher luminosities than those of the other solution classes. Furthermore, the SEDs display a systematic increase in radiative power and a shift toward higher frequency cut-offs with increasing $\delta$. This finding evidently indicates the overall increase of plasma temperature at the inner edge of the disc, as the accretion flow can extend closer to the increasingly compact horizon (see Fig. \ref{fig-08}).
        
\end{itemize}

Finally, we acknowledge that the present analysis is performed under a set of simplified assumptions. In particular, we neglect the effects of black hole rotation. We also do not take into account various dissipation mechanisms, including viscous dissipation, magnetic fields and radiative cooling processes. Needless to mention that these physical processes are important in the context of accretion physics. A fully self-consistent framework that incorporates these effects is beyond the scope of the present study. We plan to address these aspects in future work.

\section*{Data Availability}
 
The data underlying this article will be available upon reasonable request.

\section*{Acknowledgments}

Authors thank the anonymous reviewer for constructive comments and useful suggestions that helped to improve the quality of the	manuscript. Authors thank the Department of Physics, IIT Guwahati, India for providing the infrastructural support to carry out this work. SC thanks MATRICS, Science and Engineering Research Board (SERB), India, for support through grant MTR/2022/000318.

\appendix

\section{Calculation of Event Horizon} \label{eventhorizon}

\begin{widetext}

For a spherically symmetric, static spacetime, the line element is expressed as,
\begin{equation}\label{line-element}
    ds^2 = -\mathscr{F}(r)~dt^2 + \frac{1}{\mathscr{F}(r)}~dr^2 + r^2 d\Omega^2,
\end{equation}
Here, $\mathscr{F}(r)$ is a function of radial coordinate $r$. Using condition that the $g^{rr}(r_{\rm H})$ component of the metric vanishes at the event horizon, we write,
 \begin{equation}\label{rrg}
     g^{rr}(r_{\rm H}) = 0 = \mathscr{F}(r).
 \end{equation}
 Therefore, substituting $\mathscr{F}(r)$ from equation \eqref{metriccomp_dimless} into equation \eqref{rrg} and multiplying by $r^4$ on both sides, we obtain the quartic equation as,
 \begin{equation}\label{quartic}
     r^4 - 2r^3 + \delta r^2 + \eta r + \beta = 0.
 \end{equation}

In order to determine the analytical solution of quartic equation \eqref{quartic}, we first compute the solution of the following cubic equation, which is given by,
 \begin{equation}
     z^3 + m_1z^2 + m_2 z + m_3 = 0
 \end{equation}
where $m_1 = -\delta$, $m_2 = -2\eta - 4\beta$, and $m_3 = 4\delta\beta - \eta^2 - 4\beta$.

Following \cite{abramowitz1965handbook,das2001standing}, we define the following quantities as,
\begin{equation}
    \mathscr{A} = \frac{3m_2 - m_1^2}{9}, \quad \mathscr{B} = \frac{9m_1 m_2 - 27 m_3 - 2 m_1^3}{54}, \quad \mathscr{G} = \mathscr{A}^3 + \mathscr{B}^2, \quad \mathscr{C} = \sqrt{ \mathscr{B} + \sqrt[3]{\mathscr{G}}}, \quad \mathscr{D} = \sqrt{ \mathscr{B} - \sqrt[3]{\mathscr{G}}}.
\end{equation}

For $\mathscr{G}>0$, there is one real root and the other two roots are complex conjugates. In such a case, the real solution is given by,
\begin{equation}
        z_1 = (\mathscr{C} + \mathscr{D}) - \frac{m_1}{3}.
\end{equation}
On contrary, when $\mathscr{G} = 0$, all roots are real with at least two equal roots, whereas for $\mathscr{G}<0$, all roots are real and unequal. The roots are given by, 
\begin{equation}
    z_1 = 2\sqrt{-\mathscr{A}} \cos\left(\frac{\varphi}{3}\right) - \frac{m_1}{3}, \quad
    z_2 = 2\sqrt{-\mathscr{A}} \cos\left(\frac{\varphi}{3}+ 120\right) - \frac{m_1}{3}, \quad
    z_3 = 2\sqrt{-\mathscr{A}} \cos\left(\frac{\varphi}{3} + 240\right) - \frac{m_1}{3}.
\end{equation}
where $\cos \varphi = \mathscr{B}/\sqrt{-\mathscr{A}}$. We now write a quadratic equation using any one of the real solutions of the cubic equation as \cite[]{abramowitz1965handbook,das2001standing},
\begin{equation}\label{quadratic}
    w^2 + \left(-1 \pm \sqrt{1 - \delta + z_1}\right)w + \frac{1}{2}\left(z_1 \mp \sqrt{z_1^2 - 4\beta}\right) = 0.
\end{equation}
We compute the analytical solution of quartic equation \eqref{quadratic} for a given set of spacetime parameters ($\delta$, $\eta$, $\beta$). Accordingly, we obtain two real roots, which correspond to the event horizon and the Cauchy horizon, respectively.

\section{Keplerian Angular Momentum Distribution} \label{Keplerian}

For a particle to execute a circular orbit in the static, spherically symmetric spacetime described by the metric in equation \eqref{metric}, it must satisfy,
\begin{equation}\label{dphi}
    {\Phi^{\rm eff}_{\rm e}}'(r,\lambda) = \frac{\partial \Phi^{\rm eff}_{\rm e}(r,\lambda)}{\partial r}\Bigg|_{r_0,\lambda_{\rm k}} = 0,
\end{equation}
where $\Phi^{\rm eff}_{\rm e}$ represents the effective potential (see equation \eqref{diskpotential}), $\lambda_{\mathrm K}$ is the Keplerian angular momentum, and $r_0$ is the radius of the circular orbit. Using equations \eqref{diskpotential} and \eqref{dphi}, we obtain $\lambda_{\mathrm{K}}$ as a function of $r_0$ as,
\begin{equation}\label{lkep}
    \lambda^2_{\rm K}(r_0) = \frac{r^3 ~\mathscr{F}'(r)}{2\mathscr{F}^2(r)}.
\end{equation}
Here, $\mathscr{F}(r) = 1-2/r + \delta/r^2 + \eta/r^3 + \beta/r^4$, and $g_{\phi\phi} = r^2$. To determine the radius of the innermost stable circular orbit ($r_{\mathrm{I}}$) corresponding to the minimum Keplerian angular momentum ($\lambda^{\rm min}_{\mathrm{K}}$), we differentiate equation \eqref{lkep} with respect to $r_{0}$ to obtain,
\begin{equation}\label{dlkep}
    \lambda_{\rm K}'(r_0) = \frac{d\lambda_k (r_0)}{dr_0}\Bigg|_{r_{\rm I}} = 0.
\end{equation}
Using equation \eqref{lkep} in equation \eqref{dlkep}, we obtain a cubic equation given by,
\begin{equation}\label{dlkepf}
    2{r_{\rm I}}^3 \mathscr{F}(r_{\rm I})\mathscr{F}'(r_{\rm I})~^2 - {r_{\rm I}}^2\mathscr{F}(r_{\rm I})^2 \left[r_0\mathscr{F}''(r_{\rm I}) + 3\mathscr{F}'(r_{\rm I}) \right] = 0.
\end{equation}
Substituting $\mathscr{F}$, $\mathscr{F}'$ and $\mathscr{F}''$, we rewrite equation \eqref{dlkepf} in terms of $r_{\rm I}$ as,
\begin{equation}\label{r-isco}
    2{r_{\rm I}}^7 - 12{r_{\rm I}}^6 + 3(\eta+6\delta){r_{\rm I}}^5 - 4(2\delta^2+5\eta-2\beta){r_{\rm I}}^4 + 3(\beta-7\delta\eta){r_{\rm I}}^3 - 3(5\eta^2+8\delta\beta){r_{\rm I}}^2 - 37\eta\beta r_{\rm I} -24\beta^2 = 0
\end{equation}

Since the present model formalism is weakly sensitive to $\eta$ and $\beta$ parameters (see \S \ref{sec:method} for details), we ignore their contributions in equation \eqref{r-isco}. Setting $\eta = \beta = 0$ in equation \eqref{r-isco}, we have,
\begin{equation}\label{r-isco1}
    r_{\rm I}^3 - 6r_{\rm I}^2 - 9\delta r_{\rm I} - 4\delta^2 = 0.
\end{equation}
Following the methodology described in Appendix \ref{eventhorizon}, we analytically solve equation \eqref{r-isco1} to determine minimum Keplerian angular momentum ($\lambda^{\rm min}_{\rm K}$) at $r_{\rm I}$ as,
\begin{equation}
    \lambda^{\rm min}_{\rm K} = \frac{\sqrt{2(\mathscr{Y} + \mathscr{Z}+ 2\delta)} ~(\mathscr{Y} + \mathscr{Z}+ 3\delta)^2}{(\mathscr{Y} + \mathscr{Z}+ 3\delta)^2 - 2(\mathscr{Y} + \mathscr{Z}+ 3\delta) + \delta},
\end{equation}
where $\mathscr{W} = -3\delta -4, \quad \mathscr{X} = 12\delta^2 + 54\delta + 8, \quad \mathscr{U} = \mathscr{W}^3 + \mathscr{X}^2, \quad \mathscr{Y} = \sqrt{ \mathscr{X} + \sqrt[3]{\mathscr{U}}}, \quad \mathscr{Z} = \sqrt{ \mathscr{X} - \sqrt[3]{\mathscr{U}}}$.

\end{widetext}


\begin{thebibliography}{69}%
	\makeatletter
	\providecommand \@ifxundefined [1]{%
		\@ifx{#1\undefined}
	}%
	\providecommand \@ifnum [1]{%
		\ifnum #1\expandafter \@firstoftwo
		\else \expandafter \@secondoftwo
		\fi
	}%
	\providecommand \@ifx [1]{%
		\ifx #1\expandafter \@firstoftwo
		\else \expandafter \@secondoftwo
		\fi
	}%
	\providecommand \natexlab [1]{#1}%
	\providecommand \enquote  [1]{``#1''}%
	\providecommand \bibnamefont  [1]{#1}%
	\providecommand \bibfnamefont [1]{#1}%
	\providecommand \citenamefont [1]{#1}%
	\providecommand \href@noop [0]{\@secondoftwo}%
	\providecommand \href [0]{\begingroup \@sanitize@url \@href}%
	\providecommand \@href[1]{\@@startlink{#1}\@@href}%
	\providecommand \@@href[1]{\endgroup#1\@@endlink}%
	\providecommand \@sanitize@url [0]{\catcode `\\12\catcode `\$12\catcode
		`\&12\catcode `\#12\catcode `\^12\catcode `\_12\catcode `\%12\relax}%
	\providecommand \@@startlink[1]{}%
	\providecommand \@@endlink[0]{}%
	\providecommand \url  [0]{\begingroup\@sanitize@url \@url }%
	\providecommand \@url [1]{\endgroup\@href {#1}{\urlprefix }}%
	\providecommand \urlprefix  [0]{URL }%
	\providecommand \Eprint [0]{\href }%
	\providecommand \doibase [0]{http://dx.doi.org/}%
	\providecommand \selectlanguage [0]{\@gobble}%
	\providecommand \bibinfo  [0]{\@secondoftwo}%
	\providecommand \bibfield  [0]{\@secondoftwo}%
	\providecommand \translation [1]{[#1]}%
	\providecommand \BibitemOpen [0]{}%
	\providecommand \bibitemStop [0]{}%
	\providecommand \bibitemNoStop [0]{.\EOS\space}%
	\providecommand \EOS [0]{\spacefactor3000\relax}%
	\providecommand \BibitemShut  [1]{\csname bibitem#1\endcsname}%
	\let\auto@bib@innerbib\@empty
	\bibitem [{\citenamefont {{Shapiro}}\ and\ \citenamefont
		{{Teukolsky}}(1983)}]{shapiro1983black}%
	\BibitemOpen
	\bibfield  {author} {\bibinfo {author} {\bibfnamefont {S.~L.}\ \bibnamefont
			{{Shapiro}}}\ and\ \bibinfo {author} {\bibfnamefont {S.~A.}\ \bibnamefont
			{{Teukolsky}}},\ }\href {\doibase 10.1002/9783527617661} {\emph {\bibinfo
			{title} {{Black holes, white dwarfs and neutron stars. The physics of compact
					objects}}}}\ (\bibinfo {year} {1983})\BibitemShut {NoStop}%
	\bibitem [{\citenamefont {{Frank}}\ \emph {et~al.}(2002)\citenamefont
		{{Frank}}, \citenamefont {{King}},\ and\ \citenamefont
		{{Raine}}}]{frank2002accretion}%
	\BibitemOpen
	\bibfield  {author} {\bibinfo {author} {\bibfnamefont {J.}~\bibnamefont
			{{Frank}}}, \bibinfo {author} {\bibfnamefont {A.}~\bibnamefont {{King}}}, \
		and\ \bibinfo {author} {\bibfnamefont {D.~J.}\ \bibnamefont {{Raine}}},\
	}\href@noop {} {\emph {\bibinfo {title} {{Accretion Power in Astrophysics:
					Third Edition}}}}\ (\bibinfo {year} {2002})\BibitemShut {NoStop}%
	\bibitem [{\citenamefont {{Shen}}\ and\ \citenamefont
		{{Ho}}(2014)}]{shen2014diversity}%
	\BibitemOpen
	\bibfield  {author} {\bibinfo {author} {\bibfnamefont {Y.}~\bibnamefont
			{{Shen}}}\ and\ \bibinfo {author} {\bibfnamefont {L.~C.}\ \bibnamefont
			{{Ho}}},\ }\href {\doibase 10.1038/nature13712} {\bibfield  {journal}
		{\bibinfo  {journal} {\nat}\ }\textbf {\bibinfo {volume} {513}},\ \bibinfo
		{pages} {210} (\bibinfo {year} {2014})},\ \Eprint
	{http://arxiv.org/abs/1409.2887} {arXiv:1409.2887 [astro-ph.GA]} \BibitemShut
	{NoStop}%
	\bibitem [{\citenamefont {{Dexter}}\ and\ \citenamefont
		{{Agol}}(2011)}]{dexter2010quasar}%
	\BibitemOpen
	\bibfield  {author} {\bibinfo {author} {\bibfnamefont {J.}~\bibnamefont
			{{Dexter}}}\ and\ \bibinfo {author} {\bibfnamefont {E.}~\bibnamefont
			{{Agol}}},\ }\href {\doibase 10.1088/2041-8205/727/1/L24} {\bibfield
		{journal} {\bibinfo  {journal} {\apjl}\ }\textbf {\bibinfo {volume} {727}},\
		\bibinfo {eid} {L24} (\bibinfo {year} {2011})},\ \Eprint
	{http://arxiv.org/abs/1012.3169} {arXiv:1012.3169 [astro-ph.CO]} \BibitemShut
	{NoStop}%
	\bibitem [{\citenamefont {{Proga}}(2007)}]{proga2007dynamics}%
	\BibitemOpen
	\bibfield  {author} {\bibinfo {author} {\bibfnamefont {D.}~\bibnamefont
			{{Proga}}},\ }\href {\doibase 10.1086/515389} {\bibfield  {journal} {\bibinfo
			{journal} {\apj}\ }\textbf {\bibinfo {volume} {661}},\ \bibinfo {pages}
		{693} (\bibinfo {year} {2007})},\ \Eprint
	{http://arxiv.org/abs/astro-ph/0702582} {arXiv:astro-ph/0702582 [astro-ph]}
	\BibitemShut {NoStop}%
	\bibitem [{\citenamefont {{Peterson}}(1997)}]{peterson1997introduction}%
	\BibitemOpen
	\bibfield  {author} {\bibinfo {author} {\bibfnamefont {B.~M.}\ \bibnamefont
			{{Peterson}}},\ }\href@noop {} {\emph {\bibinfo {title} {{An Introduction to
					Active Galactic Nuclei}}}}\ (\bibinfo {year} {1997})\BibitemShut {NoStop}%
	\bibitem [{\citenamefont {{Fabian}}(2012)}]{fabian2012observational}%
	\BibitemOpen
	\bibfield  {author} {\bibinfo {author} {\bibfnamefont {A.~C.}\ \bibnamefont
			{{Fabian}}},\ }\href {\doibase 10.1146/annurev-astro-081811-125521}
	{\bibfield  {journal} {\bibinfo  {journal} {\araa}\ }\textbf {\bibinfo
			{volume} {50}},\ \bibinfo {pages} {455} (\bibinfo {year} {2012})},\ \Eprint
	{http://arxiv.org/abs/1204.4114} {arXiv:1204.4114 [astro-ph.CO]} \BibitemShut
	{NoStop}%
	\bibitem [{\citenamefont {{Esin}}\ \emph {et~al.}(1997)\citenamefont {{Esin}},
		\citenamefont {{McClintock}},\ and\ \citenamefont
		{{Narayan}}}]{esin1997advection}%
	\BibitemOpen
	\bibfield  {author} {\bibinfo {author} {\bibfnamefont {A.~A.}\ \bibnamefont
			{{Esin}}}, \bibinfo {author} {\bibfnamefont {J.~E.}\ \bibnamefont
			{{McClintock}}}, \ and\ \bibinfo {author} {\bibfnamefont {R.}~\bibnamefont
			{{Narayan}}},\ }\href {\doibase 10.1086/304829} {\bibfield  {journal}
		{\bibinfo  {journal} {\apj}\ }\textbf {\bibinfo {volume} {489}},\ \bibinfo
		{pages} {865} (\bibinfo {year} {1997})},\ \Eprint
	{http://arxiv.org/abs/astro-ph/9705237} {arXiv:astro-ph/9705237 [astro-ph]}
	\BibitemShut {NoStop}%
	\bibitem [{\citenamefont {{Davis}}\ \emph {et~al.}(2006)\citenamefont
		{{Davis}}, \citenamefont {{Done}},\ and\ \citenamefont
		{{Blaes}}}]{davis2006testing}%
	\BibitemOpen
	\bibfield  {author} {\bibinfo {author} {\bibfnamefont {S.~W.}\ \bibnamefont
			{{Davis}}}, \bibinfo {author} {\bibfnamefont {C.}~\bibnamefont {{Done}}}, \
		and\ \bibinfo {author} {\bibfnamefont {O.~M.}\ \bibnamefont {{Blaes}}},\
	}\href {\doibase 10.1086/505386} {\bibfield  {journal} {\bibinfo  {journal}
			{\apj}\ }\textbf {\bibinfo {volume} {647}},\ \bibinfo {pages} {525} (\bibinfo
		{year} {2006})},\ \Eprint {http://arxiv.org/abs/astro-ph/0602245}
	{arXiv:astro-ph/0602245 [astro-ph]} \BibitemShut {NoStop}%
	\bibitem [{\citenamefont {{Igumenshchev}}\ and\ \citenamefont
		{{Abramowicz}}(1999)}]{igumenshchev1999rotating}%
	\BibitemOpen
	\bibfield  {author} {\bibinfo {author} {\bibfnamefont {I.~V.}\ \bibnamefont
			{{Igumenshchev}}}\ and\ \bibinfo {author} {\bibfnamefont {M.~A.}\
			\bibnamefont {{Abramowicz}}},\ }\href {\doibase
		10.1046/j.1365-8711.1999.02220.x} {\bibfield  {journal} {\bibinfo  {journal}
			{\mnras}\ }\textbf {\bibinfo {volume} {303}},\ \bibinfo {pages} {309}
		(\bibinfo {year} {1999})}\BibitemShut {NoStop}%
	\bibitem [{\citenamefont {{Li}}\ \emph {et~al.}(2013)\citenamefont {{Li}},
		\citenamefont {{Ostriker}},\ and\ \citenamefont
		{{Sunyaev}}}]{li2013rotating}%
	\BibitemOpen
	\bibfield  {author} {\bibinfo {author} {\bibfnamefont {J.}~\bibnamefont
			{{Li}}}, \bibinfo {author} {\bibfnamefont {J.}~\bibnamefont {{Ostriker}}}, \
		and\ \bibinfo {author} {\bibfnamefont {R.}~\bibnamefont {{Sunyaev}}},\ }\href
	{\doibase 10.1088/0004-637X/767/2/105} {\bibfield  {journal} {\bibinfo
			{journal} {\apj}\ }\textbf {\bibinfo {volume} {767}},\ \bibinfo {eid} {105}
		(\bibinfo {year} {2013})},\ \Eprint {http://arxiv.org/abs/1206.4059}
	{arXiv:1206.4059 [astro-ph.GA]} \BibitemShut {NoStop}%
	\bibitem [{\citenamefont {{Yuan}}\ and\ \citenamefont
		{{Narayan}}(2014)}]{yuan2014hot}%
	\BibitemOpen
	\bibfield  {author} {\bibinfo {author} {\bibfnamefont {F.}~\bibnamefont
			{{Yuan}}}\ and\ \bibinfo {author} {\bibfnamefont {R.}~\bibnamefont
			{{Narayan}}},\ }\href {\doibase 10.1146/annurev-astro-082812-141003}
	{\bibfield  {journal} {\bibinfo  {journal} {\araa}\ }\textbf {\bibinfo
			{volume} {52}},\ \bibinfo {pages} {529} (\bibinfo {year} {2014})},\ \Eprint
	{http://arxiv.org/abs/1401.0586} {arXiv:1401.0586 [astro-ph.HE]} \BibitemShut
	{NoStop}%
	\bibitem [{\citenamefont {{Torres}}(2002)}]{torres2002accretion}%
	\BibitemOpen
	\bibfield  {author} {\bibinfo {author} {\bibfnamefont {D.~F.}\ \bibnamefont
			{{Torres}}},\ }\href {\doibase 10.1016/S0550-3213(02)00038-X} {\bibfield
		{journal} {\bibinfo  {journal} {Nuclear Physics B}\ }\textbf {\bibinfo
			{volume} {626}},\ \bibinfo {pages} {377} (\bibinfo {year} {2002})},\ \Eprint
	{http://arxiv.org/abs/hep-ph/0201154} {arXiv:hep-ph/0201154 [hep-ph]}
	\BibitemShut {NoStop}%
	\bibitem [{\citenamefont {{Guzm{\'a}n}}(2006)}]{guzman2006accretion}%
	\BibitemOpen
	\bibfield  {author} {\bibinfo {author} {\bibfnamefont {F.~S.}\ \bibnamefont
			{{Guzm{\'a}n}}},\ }\href {\doibase 10.1103/PhysRevD.73.021501} {\bibfield
		{journal} {\bibinfo  {journal} {\prd}\ }\textbf {\bibinfo {volume} {73}},\
		\bibinfo {eid} {021501} (\bibinfo {year} {2006})},\ \Eprint
	{http://arxiv.org/abs/gr-qc/0512081} {arXiv:gr-qc/0512081 [gr-qc]}
	\BibitemShut {NoStop}%
	\bibitem [{\citenamefont {{Harko}}\ \emph
		{et~al.}(2009{\natexlab{a}})\citenamefont {{Harko}}, \citenamefont
		{{Kov{\'a}cs}},\ and\ \citenamefont {{Lobo}}}]{harko2009thin}%
	\BibitemOpen
	\bibfield  {author} {\bibinfo {author} {\bibfnamefont {T.}~\bibnamefont
			{{Harko}}}, \bibinfo {author} {\bibfnamefont {Z.}~\bibnamefont
			{{Kov{\'a}cs}}}, \ and\ \bibinfo {author} {\bibfnamefont {F.~S.~N.}\
			\bibnamefont {{Lobo}}},\ }\href {\doibase 10.1103/PhysRevD.79.064001}
	{\bibfield  {journal} {\bibinfo  {journal} {\prd}\ }\textbf {\bibinfo
			{volume} {79}},\ \bibinfo {eid} {064001} (\bibinfo {year}
		{2009}{\natexlab{a}})},\ \Eprint {http://arxiv.org/abs/0901.3926}
	{arXiv:0901.3926 [gr-qc]} \BibitemShut {NoStop}%
	\bibitem [{\citenamefont {{Harko}}\ \emph
		{et~al.}(2009{\natexlab{b}})\citenamefont {{Harko}}, \citenamefont
		{{Kov{\'a}cs}},\ and\ \citenamefont {{Lobo}}}]{harko2009can}%
	\BibitemOpen
	\bibfield  {author} {\bibinfo {author} {\bibfnamefont {T.}~\bibnamefont
			{{Harko}}}, \bibinfo {author} {\bibfnamefont {Z.}~\bibnamefont
			{{Kov{\'a}cs}}}, \ and\ \bibinfo {author} {\bibfnamefont {F.~S.~N.}\
			\bibnamefont {{Lobo}}},\ }\href {\doibase 10.1088/0264-9381/26/21/215006}
	{\bibfield  {journal} {\bibinfo  {journal} {Classical and Quantum Gravity}\
		}\textbf {\bibinfo {volume} {26}},\ \bibinfo {eid} {215006} (\bibinfo {year}
		{2009}{\natexlab{b}})},\ \Eprint {http://arxiv.org/abs/0905.1355}
	{arXiv:0905.1355 [gr-qc]} \BibitemShut {NoStop}%
	\bibitem [{\citenamefont {{Kov{\'a}cs}}\ \emph {et~al.}(2009)\citenamefont
		{{Kov{\'a}cs}}, \citenamefont {{Cheng}},\ and\ \citenamefont
		{{Harko}}}]{kovacs2009can}%
	\BibitemOpen
	\bibfield  {author} {\bibinfo {author} {\bibfnamefont {Z.}~\bibnamefont
			{{Kov{\'a}cs}}}, \bibinfo {author} {\bibfnamefont {K.~S.}\ \bibnamefont
			{{Cheng}}}, \ and\ \bibinfo {author} {\bibfnamefont {T.}~\bibnamefont
			{{Harko}}},\ }\href {\doibase 10.1111/j.1365-2966.2009.15571.x} {\bibfield
		{journal} {\bibinfo  {journal} {\mnras}\ }\textbf {\bibinfo {volume} {400}},\
		\bibinfo {pages} {1632} (\bibinfo {year} {2009})},\ \Eprint
	{http://arxiv.org/abs/0908.2672} {arXiv:0908.2672 [astro-ph.HE]} \BibitemShut
	{NoStop}%
	\bibitem [{\citenamefont {{Joshi}}\ \emph {et~al.}(2014)\citenamefont
		{{Joshi}}, \citenamefont {{Malafarina}},\ and\ \citenamefont
		{{Narayan}}}]{joshi2013distinguishing}%
	\BibitemOpen
	\bibfield  {author} {\bibinfo {author} {\bibfnamefont {P.~S.}\ \bibnamefont
			{{Joshi}}}, \bibinfo {author} {\bibfnamefont {D.}~\bibnamefont
			{{Malafarina}}}, \ and\ \bibinfo {author} {\bibfnamefont {R.}~\bibnamefont
			{{Narayan}}},\ }\href {\doibase 10.1088/0264-9381/31/1/015002} {\bibfield
		{journal} {\bibinfo  {journal} {Classical and Quantum Gravity}\ }\textbf
		{\bibinfo {volume} {31}},\ \bibinfo {eid} {015002} (\bibinfo {year}
		{2014})},\ \Eprint {http://arxiv.org/abs/1304.7331} {arXiv:1304.7331 [gr-qc]}
	\BibitemShut {NoStop}%
	\bibitem [{\citenamefont {{Kov{\'a}cs}}\ and\ \citenamefont
		{{Harko}}(2010)}]{kovacs2010can}%
	\BibitemOpen
	\bibfield  {author} {\bibinfo {author} {\bibfnamefont {Z.}~\bibnamefont
			{{Kov{\'a}cs}}}\ and\ \bibinfo {author} {\bibfnamefont {T.}~\bibnamefont
			{{Harko}}},\ }\href {\doibase 10.1103/PhysRevD.82.124047} {\bibfield
		{journal} {\bibinfo  {journal} {\prd}\ }\textbf {\bibinfo {volume} {82}},\
		\bibinfo {eid} {124047} (\bibinfo {year} {2010})},\ \Eprint
	{http://arxiv.org/abs/1011.4127} {arXiv:1011.4127 [gr-qc]} \BibitemShut
	{NoStop}%
	\bibitem [{\citenamefont {{Harko}}\ \emph {et~al.}(2010)\citenamefont
		{{Harko}}, \citenamefont {{Kov{\'a}cs}},\ and\ \citenamefont
		{{Lobo}}}]{harko2010thin}%
	\BibitemOpen
	\bibfield  {author} {\bibinfo {author} {\bibfnamefont {T.}~\bibnamefont
			{{Harko}}}, \bibinfo {author} {\bibfnamefont {Z.}~\bibnamefont
			{{Kov{\'a}cs}}}, \ and\ \bibinfo {author} {\bibfnamefont {F.~S.~N.}\
			\bibnamefont {{Lobo}}},\ }\href {\doibase 10.1088/0264-9381/27/10/105010}
	{\bibfield  {journal} {\bibinfo  {journal} {Classical and Quantum Gravity}\
		}\textbf {\bibinfo {volume} {27}},\ \bibinfo {eid} {105010} (\bibinfo {year}
		{2010})},\ \Eprint {http://arxiv.org/abs/0909.1267} {arXiv:0909.1267 [gr-qc]}
	\BibitemShut {NoStop}%
	\bibitem [{\citenamefont {{Pun}}\ \emph {et~al.}(2008)\citenamefont {{Pun}},
		\citenamefont {{Kov{\'a}cs}},\ and\ \citenamefont {{Harko}}}]{pun2008thin}%
	\BibitemOpen
	\bibfield  {author} {\bibinfo {author} {\bibfnamefont {C.~S.~J.}\
			\bibnamefont {{Pun}}}, \bibinfo {author} {\bibfnamefont {Z.}~\bibnamefont
			{{Kov{\'a}cs}}}, \ and\ \bibinfo {author} {\bibfnamefont {T.}~\bibnamefont
			{{Harko}}},\ }\href {\doibase 10.1103/PhysRevD.78.084015} {\bibfield
		{journal} {\bibinfo  {journal} {\prd}\ }\textbf {\bibinfo {volume} {78}},\
		\bibinfo {eid} {084015} (\bibinfo {year} {2008})},\ \Eprint
	{http://arxiv.org/abs/0809.1284} {arXiv:0809.1284 [gr-qc]} \BibitemShut
	{NoStop}%
	\bibitem [{\citenamefont {{Heydari-Fard}}(2010)}]{heydari2010black}%
	\BibitemOpen
	\bibfield  {author} {\bibinfo {author} {\bibfnamefont {M.}~\bibnamefont
			{{Heydari-Fard}}},\ }\href {\doibase 10.1088/0264-9381/27/23/235004}
	{\bibfield  {journal} {\bibinfo  {journal} {Classical and Quantum Gravity}\
		}\textbf {\bibinfo {volume} {27}},\ \bibinfo {eid} {235004} (\bibinfo {year}
		{2010})}\BibitemShut {NoStop}%
	\bibitem [{\citenamefont {{Harko}}\ \emph {et~al.}(2011)\citenamefont
		{{Harko}}, \citenamefont {{Kov{\'a}cs}},\ and\ \citenamefont
		{{Lobo}}}]{harko2011thin}%
	\BibitemOpen
	\bibfield  {author} {\bibinfo {author} {\bibfnamefont {T.}~\bibnamefont
			{{Harko}}}, \bibinfo {author} {\bibfnamefont {Z.}~\bibnamefont
			{{Kov{\'a}cs}}}, \ and\ \bibinfo {author} {\bibfnamefont {F.~S.~N.}\
			\bibnamefont {{Lobo}}},\ }\href {\doibase 10.1088/0264-9381/28/16/165001}
	{\bibfield  {journal} {\bibinfo  {journal} {Classical and Quantum Gravity}\
		}\textbf {\bibinfo {volume} {28}},\ \bibinfo {eid} {165001} (\bibinfo {year}
		{2011})},\ \Eprint {http://arxiv.org/abs/1009.1958} {arXiv:1009.1958 [gr-qc]}
	\BibitemShut {NoStop}%
	\bibitem [{\citenamefont {{Harko}}\ \emph
		{et~al.}(2009{\natexlab{c}})\citenamefont {{Harko}}, \citenamefont
		{{Kov{\'a}cs}},\ and\ \citenamefont {{Lobo}}}]{harko2009testing}%
	\BibitemOpen
	\bibfield  {author} {\bibinfo {author} {\bibfnamefont {T.}~\bibnamefont
			{{Harko}}}, \bibinfo {author} {\bibfnamefont {Z.}~\bibnamefont
			{{Kov{\'a}cs}}}, \ and\ \bibinfo {author} {\bibfnamefont {F.~S.~N.}\
			\bibnamefont {{Lobo}}},\ }\href {\doibase 10.1103/PhysRevD.80.044021}
	{\bibfield  {journal} {\bibinfo  {journal} {\prd}\ }\textbf {\bibinfo
			{volume} {80}},\ \bibinfo {eid} {044021} (\bibinfo {year}
		{2009}{\natexlab{c}})},\ \Eprint {http://arxiv.org/abs/0907.1449}
	{arXiv:0907.1449 [gr-qc]} \BibitemShut {NoStop}%
	\bibitem [{\citenamefont {{Uniyal}}\ \emph {et~al.}(2024)\citenamefont
		{{Uniyal}}, \citenamefont {{Chakrabarti}},\ and\ \citenamefont
		{{Das}}}]{uniyal2024}%
	\BibitemOpen
	\bibfield  {author} {\bibinfo {author} {\bibfnamefont {A.}~\bibnamefont
			{{Uniyal}}}, \bibinfo {author} {\bibfnamefont {S.}~\bibnamefont
			{{Chakrabarti}}}, \ and\ \bibinfo {author} {\bibfnamefont {S.}~\bibnamefont
			{{Das}}},\ }\href {\doibase 10.1016/j.dark.2024.101429} {\bibfield  {journal}
		{\bibinfo  {journal} {Physics of the Dark Universe}\ }\textbf {\bibinfo
			{volume} {44}},\ \bibinfo {eid} {101429} (\bibinfo {year} {2024})},\ \Eprint
	{http://arxiv.org/abs/2306.14434} {arXiv:2306.14434 [astro-ph.HE]}
	\BibitemShut {NoStop}%
	\bibitem [{\citenamefont {{Will}}(2018)}]{will2018theory}%
	\BibitemOpen
	\bibfield  {author} {\bibinfo {author} {\bibfnamefont {C.~M.}\ \bibnamefont
			{{Will}}},\ }\href {\doibase 10.1017/9781316338612} {\emph {\bibinfo {title}
			{{Theory and Experiment in Gravitational Physics}}}}\ (\bibinfo {year}
	{2018})\BibitemShut {NoStop}%
	\bibitem [{\citenamefont {{Abbott}}\ \emph {et~al.}(2016)\citenamefont
		{{Abbott}}, \citenamefont {{Abbott}}, \citenamefont {{Abbott}}, \citenamefont
		{{Abernathy}}, \citenamefont {{Acernese}}, \citenamefont {{Ackley}},
		\citenamefont {{Adams}}, \citenamefont {{Adams}}, \citenamefont {{Addesso}},
		\citenamefont {{Adhikari}}, \citenamefont {{Adya}}, \citenamefont
		{{Affeldt}}, \citenamefont {{Agathos}}, \citenamefont {{Agatsuma}},
		\citenamefont {{Aggarwal}}, \citenamefont {{Aguiar}}, \citenamefont
		{{Aiello}}, \citenamefont {{Ain}}, \citenamefont {{Ajith}}, \citenamefont
		{{Allen}}, \citenamefont {{Allocca}}, \citenamefont {{Altin}}, \citenamefont
		{{Anderson}}, \citenamefont {{Anderson}}, \citenamefont {{Arai}},
		\citenamefont {{Araya}}, \citenamefont {{Arceneaux}}, \citenamefont
		{{Areeda}}, \citenamefont {{Arnaud}}, \citenamefont {{Arun}}, \citenamefont
		{{Ascenzi}}, \citenamefont {{Ashton}}, \citenamefont {{Ast}}, \citenamefont
		{{Aston}}, \citenamefont {{Astone}}, \citenamefont {{Aufmuth}}, \citenamefont
		{{Aulbert}}, \citenamefont {{Babak}}, \citenamefont {{Bacon}}, \citenamefont
		{{Bader}}, \citenamefont {{Baker}}, \citenamefont {{Baldaccini}},
		\citenamefont {{Ballardin}}, \citenamefont {{Ballmer}}, \citenamefont
		{{Barayoga}}, \citenamefont {{Barclay}}, \citenamefont {{Barish}},
		\citenamefont {{Barker}}, \citenamefont {{Barone}}, \citenamefont {{Barr}},
		\citenamefont {{Barsotti}}, \citenamefont {{Barsuglia}}, \citenamefont
		{{Barta}}, \citenamefont {{Bartlett}}, \citenamefont {{Bartos}},
		\citenamefont {{Bassiri}}, \citenamefont {{Basti}}, \citenamefont {{Batch}},
		\citenamefont {{Baune}}, \citenamefont {{Bavigadda}}, \citenamefont
		{{Bazzan}}, \citenamefont {{Behnke}}, \citenamefont {{Bejger}}, \citenamefont
		{{Bell}}, \citenamefont {{Bell}}, \citenamefont {{Berger}}, \citenamefont
		{{Bergman}}, \citenamefont {{Bergmann}}, \citenamefont {{Berry}},
		\citenamefont {{Bersanetti}}, \citenamefont {{Bertolini}}, \citenamefont
		{{Betzwieser}}, \citenamefont {{Bhagwat}}, \citenamefont {{Bhandare}},
		\citenamefont {{Bilenko}}, \citenamefont {{Billingsley}}, \citenamefont
		{{Birch}}, \citenamefont {{Birney}}, \citenamefont {{Biscans}}, \citenamefont
		{{Bisht}}, \citenamefont {{Bitossi}}, \citenamefont {{Biwer}}, \citenamefont
		{{Bizouard}}, \citenamefont {{Blackburn}}, \citenamefont {{Blair}},
		\citenamefont {{Blair}}, \citenamefont {{Blair}}, \citenamefont {{Bloemen}},
		\citenamefont {{Bock}}, \citenamefont {{Bodiya}}, \citenamefont {{Boer}},
		\citenamefont {{Bogaert}}, \citenamefont {{Bogan}}, \citenamefont {{Bohe}},
		\citenamefont {{Bojtos}}, \citenamefont {{Bond}}, \citenamefont {{Bondu}},
		\citenamefont {{Bonnand}}, \citenamefont {{Boom}}, \citenamefont {{Bork}},
		\citenamefont {{Boschi}}, \citenamefont {{Bose}}, \citenamefont
		{{Bouffanais}}, \citenamefont {{Bozzi}}, \citenamefont {{Bradaschia}},
		\citenamefont {{Brady}}, \citenamefont {{Braginsky}}, \citenamefont
		{{Branchesi}}, \citenamefont {{Brau}}, \citenamefont {{Briant}},
		\citenamefont {{Brillet}}, \citenamefont {{Brinkmann}}, \citenamefont
		{{Brisson}}, \citenamefont {{Brockill}}, \citenamefont {{Brooks}},
		\citenamefont {{Brown}}, \citenamefont {{Brown}}, \citenamefont {{Brown}},
		\citenamefont {{Buchanan}}, \citenamefont {{Buikema}}, \citenamefont
		{{Bulik}}, \citenamefont {{Bulten}}, \citenamefont {{Buonanno}},
		\citenamefont {{Buskulic}}, \citenamefont {{Buy}}, \citenamefont {{Byer}},
		\citenamefont {{Cadonati}}, \citenamefont {{Cagnoli}}, \citenamefont
		{{Cahillane}}, \citenamefont {{Calder{\'o}n Bustillo}}, \citenamefont
		{{Callister}}, \citenamefont {{Calloni}}, \citenamefont {{Camp}},
		\citenamefont {{Cannon}}, \citenamefont {{Cao}}, \citenamefont {{Capano}},
		\citenamefont {{Capocasa}}, \citenamefont {{Carbognani}}, \citenamefont
		{{Caride}}, \citenamefont {{Casanueva Diaz}}, \citenamefont {{Casentini}},
		\citenamefont {{Caudill}}, \citenamefont {{Cavagli{\`a}}}, \citenamefont
		{{Cavalier}}, \citenamefont {{Cavalieri}}, \citenamefont {{Cella}},
		\citenamefont {{Cepeda}}, \citenamefont {{Cerboni Baiardi}}, \citenamefont
		{{Cerretani}}, \citenamefont {{Cesarini}}, \citenamefont {{Chakraborty}},
		\citenamefont {{Chalermsongsak}}, \citenamefont {{Chamberlin}}, \citenamefont
		{{Chan}}, \citenamefont {{Chao}}, \citenamefont {{Charlton}}, \citenamefont
		{{Chassande-Mottin}}, \citenamefont {{Chen}}, \citenamefont {{Chen}},
		\citenamefont {{Cheng}}, \citenamefont {{Chincarini}}, \citenamefont
		{{Chiummo}}, \citenamefont {{Cho}}, \citenamefont {{Cho}}, \citenamefont
		{{Chow}}, \citenamefont {{Christensen}}, \citenamefont {{Chu}}, \citenamefont
		{{Chua}}, \citenamefont {{Chung}}, \citenamefont {{Ciani}}, \citenamefont
		{{Clara}}, \citenamefont {{Clark}}, \citenamefont {{Cleva}}, \citenamefont
		{{Coccia}}, \citenamefont {{Cohadon}}, \citenamefont {{Colla}}, \citenamefont
		{{Collette}}, \citenamefont {{Cominsky}}, \citenamefont {{Constancio}},
		\citenamefont {{Conte}}, \citenamefont {{Conti}}, \citenamefont {{Cook}},
		\citenamefont {{Corbitt}}, \citenamefont {{Cornish}}, \citenamefont
		{{Corsi}}, \citenamefont {{Cortese}}, \citenamefont {{Costa}}, \citenamefont
		{{Coughlin}}, \citenamefont {{Coughlin}}, \citenamefont {{Coulon}},
		\citenamefont {{Countryman}}, \citenamefont {{Couvares}}, \citenamefont
		{{Cowan}}, \citenamefont {{Coward}}, \citenamefont {{Cowart}}, \citenamefont
		{{Coyne}}, \citenamefont {{Coyne}}, \citenamefont {{Craig}}, \citenamefont
		{{Creighton}},\ and\ \citenamefont {{Cripe}}}]{abbott2016gw150914}%
	\BibitemOpen
	\bibfield  {author} {\bibinfo {author} {\bibfnamefont {B.~P.}\ \bibnamefont
			{{Abbott}}}, \bibinfo {author} {\bibfnamefont {R.}~\bibnamefont {{Abbott}}},
		\bibinfo {author} {\bibfnamefont {T.~D.}\ \bibnamefont {{Abbott}}}, \bibinfo
		{author} {\bibfnamefont {M.~R.}\ \bibnamefont {{Abernathy}}}, \bibinfo
		{author} {\bibfnamefont {F.}~\bibnamefont {{Acernese}}}, \bibinfo {author}
		{\bibfnamefont {K.}~\bibnamefont {{Ackley}}}, \bibinfo {author}
		{\bibfnamefont {C.}~\bibnamefont {{Adams}}}, \bibinfo {author} {\bibfnamefont
			{T.}~\bibnamefont {{Adams}}}, \bibinfo {author} {\bibfnamefont
			{P.}~\bibnamefont {{Addesso}}}, \bibinfo {author} {\bibfnamefont {R.~X.}\
			\bibnamefont {{Adhikari}}}, \bibinfo {author} {\bibfnamefont {V.~B.}\
			\bibnamefont {{Adya}}}, \bibinfo {author} {\bibfnamefont {C.}~\bibnamefont
			{{Affeldt}}}, \bibinfo {author} {\bibfnamefont {M.}~\bibnamefont
			{{Agathos}}}, \bibinfo {author} {\bibfnamefont {K.}~\bibnamefont
			{{Agatsuma}}}, \bibinfo {author} {\bibfnamefont {N.}~\bibnamefont
			{{Aggarwal}}}, \bibinfo {author} {\bibfnamefont {O.~D.}\ \bibnamefont
			{{Aguiar}}}, \bibinfo {author} {\bibfnamefont {L.}~\bibnamefont {{Aiello}}},
		\bibinfo {author} {\bibfnamefont {A.}~\bibnamefont {{Ain}}}, \bibinfo
		{author} {\bibfnamefont {P.}~\bibnamefont {{Ajith}}}, \bibinfo {author}
		{\bibfnamefont {B.}~\bibnamefont {{Allen}}}, \bibinfo {author} {\bibfnamefont
			{A.}~\bibnamefont {{Allocca}}}, \bibinfo {author} {\bibfnamefont {P.~A.}\
			\bibnamefont {{Altin}}}, \bibinfo {author} {\bibfnamefont {S.~B.}\
			\bibnamefont {{Anderson}}}, \bibinfo {author} {\bibfnamefont {W.~G.}\
			\bibnamefont {{Anderson}}}, \bibinfo {author} {\bibfnamefont
			{K.}~\bibnamefont {{Arai}}}, \bibinfo {author} {\bibfnamefont {M.~C.}\
			\bibnamefont {{Araya}}}, \bibinfo {author} {\bibfnamefont {C.~C.}\
			\bibnamefont {{Arceneaux}}}, \bibinfo {author} {\bibfnamefont {J.~S.}\
			\bibnamefont {{Areeda}}}, \bibinfo {author} {\bibfnamefont {N.}~\bibnamefont
			{{Arnaud}}}, \bibinfo {author} {\bibfnamefont {K.~G.}\ \bibnamefont
			{{Arun}}}, \bibinfo {author} {\bibfnamefont {S.}~\bibnamefont {{Ascenzi}}},
		\bibinfo {author} {\bibfnamefont {G.}~\bibnamefont {{Ashton}}}, \bibinfo
		{author} {\bibfnamefont {M.}~\bibnamefont {{Ast}}}, \bibinfo {author}
		{\bibfnamefont {S.~M.}\ \bibnamefont {{Aston}}}, \bibinfo {author}
		{\bibfnamefont {P.}~\bibnamefont {{Astone}}}, \bibinfo {author}
		{\bibfnamefont {P.}~\bibnamefont {{Aufmuth}}}, \bibinfo {author}
		{\bibfnamefont {C.}~\bibnamefont {{Aulbert}}}, \bibinfo {author}
		{\bibfnamefont {S.}~\bibnamefont {{Babak}}}, \bibinfo {author} {\bibfnamefont
			{P.}~\bibnamefont {{Bacon}}}, \bibinfo {author} {\bibfnamefont {M.~K.~M.}\
			\bibnamefont {{Bader}}}, \bibinfo {author} {\bibfnamefont {P.~T.}\
			\bibnamefont {{Baker}}}, \bibinfo {author} {\bibfnamefont {F.}~\bibnamefont
			{{Baldaccini}}}, \bibinfo {author} {\bibfnamefont {G.}~\bibnamefont
			{{Ballardin}}}, \bibinfo {author} {\bibfnamefont {S.~W.}\ \bibnamefont
			{{Ballmer}}}, \bibinfo {author} {\bibfnamefont {J.~C.}\ \bibnamefont
			{{Barayoga}}}, \bibinfo {author} {\bibfnamefont {S.~E.}\ \bibnamefont
			{{Barclay}}}, \bibinfo {author} {\bibfnamefont {B.~C.}\ \bibnamefont
			{{Barish}}}, \bibinfo {author} {\bibfnamefont {D.}~\bibnamefont {{Barker}}},
		\bibinfo {author} {\bibfnamefont {F.}~\bibnamefont {{Barone}}}, \bibinfo
		{author} {\bibfnamefont {B.}~\bibnamefont {{Barr}}}, \bibinfo {author}
		{\bibfnamefont {L.}~\bibnamefont {{Barsotti}}}, \bibinfo {author}
		{\bibfnamefont {M.}~\bibnamefont {{Barsuglia}}}, \bibinfo {author}
		{\bibfnamefont {D.}~\bibnamefont {{Barta}}}, \bibinfo {author} {\bibfnamefont
			{J.}~\bibnamefont {{Bartlett}}}, \bibinfo {author} {\bibfnamefont
			{I.}~\bibnamefont {{Bartos}}}, \bibinfo {author} {\bibfnamefont
			{R.}~\bibnamefont {{Bassiri}}}, \bibinfo {author} {\bibfnamefont
			{A.}~\bibnamefont {{Basti}}}, \bibinfo {author} {\bibfnamefont {J.~C.}\
			\bibnamefont {{Batch}}}, \bibinfo {author} {\bibfnamefont {C.}~\bibnamefont
			{{Baune}}}, \bibinfo {author} {\bibfnamefont {V.}~\bibnamefont
			{{Bavigadda}}}, \bibinfo {author} {\bibfnamefont {M.}~\bibnamefont
			{{Bazzan}}}, \bibinfo {author} {\bibfnamefont {B.}~\bibnamefont {{Behnke}}},
		\bibinfo {author} {\bibfnamefont {M.}~\bibnamefont {{Bejger}}}, \bibinfo
		{author} {\bibfnamefont {A.~S.}\ \bibnamefont {{Bell}}}, \bibinfo {author}
		{\bibfnamefont {C.~J.}\ \bibnamefont {{Bell}}}, \bibinfo {author}
		{\bibfnamefont {B.~K.}\ \bibnamefont {{Berger}}}, \bibinfo {author}
		{\bibfnamefont {J.}~\bibnamefont {{Bergman}}}, \bibinfo {author}
		{\bibfnamefont {G.}~\bibnamefont {{Bergmann}}}, \bibinfo {author}
		{\bibfnamefont {C.~P.~L.}\ \bibnamefont {{Berry}}}, \bibinfo {author}
		{\bibfnamefont {D.}~\bibnamefont {{Bersanetti}}}, \bibinfo {author}
		{\bibfnamefont {A.}~\bibnamefont {{Bertolini}}}, \bibinfo {author}
		{\bibfnamefont {J.}~\bibnamefont {{Betzwieser}}}, \bibinfo {author}
		{\bibfnamefont {S.}~\bibnamefont {{Bhagwat}}}, \bibinfo {author}
		{\bibfnamefont {R.}~\bibnamefont {{Bhandare}}}, \bibinfo {author}
		{\bibfnamefont {I.~A.}\ \bibnamefont {{Bilenko}}}, \bibinfo {author}
		{\bibfnamefont {G.}~\bibnamefont {{Billingsley}}}, \bibinfo {author}
		{\bibfnamefont {J.}~\bibnamefont {{Birch}}}, \bibinfo {author} {\bibfnamefont
			{I.~A.}\ \bibnamefont {{Birney}}}, \bibinfo {author} {\bibfnamefont
			{S.}~\bibnamefont {{Biscans}}}, \bibinfo {author} {\bibfnamefont
			{A.}~\bibnamefont {{Bisht}}}, \bibinfo {author} {\bibfnamefont
			{M.}~\bibnamefont {{Bitossi}}}, \bibinfo {author} {\bibfnamefont
			{C.}~\bibnamefont {{Biwer}}}, \bibinfo {author} {\bibfnamefont {M.~A.}\
			\bibnamefont {{Bizouard}}}, \bibinfo {author} {\bibfnamefont {J.~K.}\
			\bibnamefont {{Blackburn}}}, \bibinfo {author} {\bibfnamefont {C.~D.}\
			\bibnamefont {{Blair}}}, \bibinfo {author} {\bibfnamefont {D.~G.}\
			\bibnamefont {{Blair}}}, \bibinfo {author} {\bibfnamefont {R.~M.}\
			\bibnamefont {{Blair}}}, \bibinfo {author} {\bibfnamefont {S.}~\bibnamefont
			{{Bloemen}}}, \bibinfo {author} {\bibfnamefont {O.}~\bibnamefont {{Bock}}},
		\bibinfo {author} {\bibfnamefont {T.~P.}\ \bibnamefont {{Bodiya}}}, \bibinfo
		{author} {\bibfnamefont {M.}~\bibnamefont {{Boer}}}, \bibinfo {author}
		{\bibfnamefont {G.}~\bibnamefont {{Bogaert}}}, \bibinfo {author}
		{\bibfnamefont {C.}~\bibnamefont {{Bogan}}}, \bibinfo {author} {\bibfnamefont
			{A.}~\bibnamefont {{Bohe}}}, \bibinfo {author} {\bibfnamefont
			{P.}~\bibnamefont {{Bojtos}}}, \bibinfo {author} {\bibfnamefont
			{C.}~\bibnamefont {{Bond}}}, \bibinfo {author} {\bibfnamefont
			{F.}~\bibnamefont {{Bondu}}}, \bibinfo {author} {\bibfnamefont
			{R.}~\bibnamefont {{Bonnand}}}, \bibinfo {author} {\bibfnamefont {B.~A.}\
			\bibnamefont {{Boom}}}, \bibinfo {author} {\bibfnamefont {R.}~\bibnamefont
			{{Bork}}}, \bibinfo {author} {\bibfnamefont {V.}~\bibnamefont {{Boschi}}},
		\bibinfo {author} {\bibfnamefont {S.}~\bibnamefont {{Bose}}}, \bibinfo
		{author} {\bibfnamefont {Y.}~\bibnamefont {{Bouffanais}}}, \bibinfo {author}
		{\bibfnamefont {A.}~\bibnamefont {{Bozzi}}}, \bibinfo {author} {\bibfnamefont
			{C.}~\bibnamefont {{Bradaschia}}}, \bibinfo {author} {\bibfnamefont {P.~R.}\
			\bibnamefont {{Brady}}}, \bibinfo {author} {\bibfnamefont {V.~B.}\
			\bibnamefont {{Braginsky}}}, \bibinfo {author} {\bibfnamefont
			{M.}~\bibnamefont {{Branchesi}}}, \bibinfo {author} {\bibfnamefont {J.~E.}\
			\bibnamefont {{Brau}}}, \bibinfo {author} {\bibfnamefont {T.}~\bibnamefont
			{{Briant}}}, \bibinfo {author} {\bibfnamefont {A.}~\bibnamefont {{Brillet}}},
		\bibinfo {author} {\bibfnamefont {M.}~\bibnamefont {{Brinkmann}}}, \bibinfo
		{author} {\bibfnamefont {V.}~\bibnamefont {{Brisson}}}, \bibinfo {author}
		{\bibfnamefont {P.}~\bibnamefont {{Brockill}}}, \bibinfo {author}
		{\bibfnamefont {A.~F.}\ \bibnamefont {{Brooks}}}, \bibinfo {author}
		{\bibfnamefont {D.~A.}\ \bibnamefont {{Brown}}}, \bibinfo {author}
		{\bibfnamefont {D.~D.}\ \bibnamefont {{Brown}}}, \bibinfo {author}
		{\bibfnamefont {N.~M.}\ \bibnamefont {{Brown}}}, \bibinfo {author}
		{\bibfnamefont {C.~C.}\ \bibnamefont {{Buchanan}}}, \bibinfo {author}
		{\bibfnamefont {A.}~\bibnamefont {{Buikema}}}, \bibinfo {author}
		{\bibfnamefont {T.}~\bibnamefont {{Bulik}}}, \bibinfo {author} {\bibfnamefont
			{H.~J.}\ \bibnamefont {{Bulten}}}, \bibinfo {author} {\bibfnamefont
			{A.}~\bibnamefont {{Buonanno}}}, \bibinfo {author} {\bibfnamefont
			{D.}~\bibnamefont {{Buskulic}}}, \bibinfo {author} {\bibfnamefont
			{C.}~\bibnamefont {{Buy}}}, \bibinfo {author} {\bibfnamefont {R.~L.}\
			\bibnamefont {{Byer}}}, \bibinfo {author} {\bibfnamefont {L.}~\bibnamefont
			{{Cadonati}}}, \bibinfo {author} {\bibfnamefont {G.}~\bibnamefont
			{{Cagnoli}}}, \bibinfo {author} {\bibfnamefont {C.}~\bibnamefont
			{{Cahillane}}}, \bibinfo {author} {\bibfnamefont {J.}~\bibnamefont
			{{Calder{\'o}n Bustillo}}}, \bibinfo {author} {\bibfnamefont
			{T.}~\bibnamefont {{Callister}}}, \bibinfo {author} {\bibfnamefont
			{E.}~\bibnamefont {{Calloni}}}, \bibinfo {author} {\bibfnamefont {J.~B.}\
			\bibnamefont {{Camp}}}, \bibinfo {author} {\bibfnamefont {K.~C.}\
			\bibnamefont {{Cannon}}}, \bibinfo {author} {\bibfnamefont {J.}~\bibnamefont
			{{Cao}}}, \bibinfo {author} {\bibfnamefont {C.~D.}\ \bibnamefont {{Capano}}},
		\bibinfo {author} {\bibfnamefont {E.}~\bibnamefont {{Capocasa}}}, \bibinfo
		{author} {\bibfnamefont {F.}~\bibnamefont {{Carbognani}}}, \bibinfo {author}
		{\bibfnamefont {S.}~\bibnamefont {{Caride}}}, \bibinfo {author}
		{\bibfnamefont {J.}~\bibnamefont {{Casanueva Diaz}}}, \bibinfo {author}
		{\bibfnamefont {C.}~\bibnamefont {{Casentini}}}, \bibinfo {author}
		{\bibfnamefont {S.}~\bibnamefont {{Caudill}}}, \bibinfo {author}
		{\bibfnamefont {M.}~\bibnamefont {{Cavagli{\`a}}}}, \bibinfo {author}
		{\bibfnamefont {F.}~\bibnamefont {{Cavalier}}}, \bibinfo {author}
		{\bibfnamefont {R.}~\bibnamefont {{Cavalieri}}}, \bibinfo {author}
		{\bibfnamefont {G.}~\bibnamefont {{Cella}}}, \bibinfo {author} {\bibfnamefont
			{C.~B.}\ \bibnamefont {{Cepeda}}}, \bibinfo {author} {\bibfnamefont
			{L.}~\bibnamefont {{Cerboni Baiardi}}}, \bibinfo {author} {\bibfnamefont
			{G.}~\bibnamefont {{Cerretani}}}, \bibinfo {author} {\bibfnamefont
			{E.}~\bibnamefont {{Cesarini}}}, \bibinfo {author} {\bibfnamefont
			{R.}~\bibnamefont {{Chakraborty}}}, \bibinfo {author} {\bibfnamefont
			{T.}~\bibnamefont {{Chalermsongsak}}}, \bibinfo {author} {\bibfnamefont
			{S.~J.}\ \bibnamefont {{Chamberlin}}}, \bibinfo {author} {\bibfnamefont
			{M.}~\bibnamefont {{Chan}}}, \bibinfo {author} {\bibfnamefont
			{S.}~\bibnamefont {{Chao}}}, \bibinfo {author} {\bibfnamefont
			{P.}~\bibnamefont {{Charlton}}}, \bibinfo {author} {\bibfnamefont
			{E.}~\bibnamefont {{Chassande-Mottin}}}, \bibinfo {author} {\bibfnamefont
			{H.~Y.}\ \bibnamefont {{Chen}}}, \bibinfo {author} {\bibfnamefont
			{Y.}~\bibnamefont {{Chen}}}, \bibinfo {author} {\bibfnamefont
			{C.}~\bibnamefont {{Cheng}}}, \bibinfo {author} {\bibfnamefont
			{A.}~\bibnamefont {{Chincarini}}}, \bibinfo {author} {\bibfnamefont
			{A.}~\bibnamefont {{Chiummo}}}, \bibinfo {author} {\bibfnamefont {H.~S.}\
			\bibnamefont {{Cho}}}, \bibinfo {author} {\bibfnamefont {M.}~\bibnamefont
			{{Cho}}}, \bibinfo {author} {\bibfnamefont {J.~H.}\ \bibnamefont {{Chow}}},
		\bibinfo {author} {\bibfnamefont {N.}~\bibnamefont {{Christensen}}}, \bibinfo
		{author} {\bibfnamefont {Q.}~\bibnamefont {{Chu}}}, \bibinfo {author}
		{\bibfnamefont {S.}~\bibnamefont {{Chua}}}, \bibinfo {author} {\bibfnamefont
			{S.}~\bibnamefont {{Chung}}}, \bibinfo {author} {\bibfnamefont
			{G.}~\bibnamefont {{Ciani}}}, \bibinfo {author} {\bibfnamefont
			{F.}~\bibnamefont {{Clara}}}, \bibinfo {author} {\bibfnamefont {J.~A.}\
			\bibnamefont {{Clark}}}, \bibinfo {author} {\bibfnamefont {F.}~\bibnamefont
			{{Cleva}}}, \bibinfo {author} {\bibfnamefont {E.}~\bibnamefont {{Coccia}}},
		\bibinfo {author} {\bibfnamefont {P.-F.}\ \bibnamefont {{Cohadon}}}, \bibinfo
		{author} {\bibfnamefont {A.}~\bibnamefont {{Colla}}}, \bibinfo {author}
		{\bibfnamefont {C.~G.}\ \bibnamefont {{Collette}}}, \bibinfo {author}
		{\bibfnamefont {L.}~\bibnamefont {{Cominsky}}}, \bibinfo {author}
		{\bibfnamefont {M.}~\bibnamefont {{Constancio}}}, \bibinfo {author}
		{\bibfnamefont {A.}~\bibnamefont {{Conte}}}, \bibinfo {author} {\bibfnamefont
			{L.}~\bibnamefont {{Conti}}}, \bibinfo {author} {\bibfnamefont
			{D.}~\bibnamefont {{Cook}}}, \bibinfo {author} {\bibfnamefont {T.~R.}\
			\bibnamefont {{Corbitt}}}, \bibinfo {author} {\bibfnamefont {N.}~\bibnamefont
			{{Cornish}}}, \bibinfo {author} {\bibfnamefont {A.}~\bibnamefont {{Corsi}}},
		\bibinfo {author} {\bibfnamefont {S.}~\bibnamefont {{Cortese}}}, \bibinfo
		{author} {\bibfnamefont {C.~A.}\ \bibnamefont {{Costa}}}, \bibinfo {author}
		{\bibfnamefont {M.~W.}\ \bibnamefont {{Coughlin}}}, \bibinfo {author}
		{\bibfnamefont {S.~B.}\ \bibnamefont {{Coughlin}}}, \bibinfo {author}
		{\bibfnamefont {J.-P.}\ \bibnamefont {{Coulon}}}, \bibinfo {author}
		{\bibfnamefont {S.~T.}\ \bibnamefont {{Countryman}}}, \bibinfo {author}
		{\bibfnamefont {P.}~\bibnamefont {{Couvares}}}, \bibinfo {author}
		{\bibfnamefont {E.~E.}\ \bibnamefont {{Cowan}}}, \bibinfo {author}
		{\bibfnamefont {D.~M.}\ \bibnamefont {{Coward}}}, \bibinfo {author}
		{\bibfnamefont {M.~J.}\ \bibnamefont {{Cowart}}}, \bibinfo {author}
		{\bibfnamefont {D.~C.}\ \bibnamefont {{Coyne}}}, \bibinfo {author}
		{\bibfnamefont {R.}~\bibnamefont {{Coyne}}}, \bibinfo {author} {\bibfnamefont
			{K.}~\bibnamefont {{Craig}}}, \bibinfo {author} {\bibfnamefont {J.~D.~E.}\
			\bibnamefont {{Creighton}}}, \ and\ \bibinfo {author} {\bibfnamefont
			{J.}~\bibnamefont {{Cripe}}},\ }\href {\doibase
		10.1103/PhysRevLett.116.131103} {\bibfield  {journal} {\bibinfo  {journal}
			{\prl}\ }\textbf {\bibinfo {volume} {116}},\ \bibinfo {eid} {131103}
		(\bibinfo {year} {2016})},\ \Eprint {http://arxiv.org/abs/1602.03838}
	{arXiv:1602.03838 [gr-qc]} \BibitemShut {NoStop}%
	\bibitem [{\citenamefont {{Clifton}}\ \emph {et~al.}(2012)\citenamefont
		{{Clifton}}, \citenamefont {{Ferreira}}, \citenamefont {{Padilla}},\ and\
		\citenamefont {{Skordis}}}]{clifton2012modified}%
	\BibitemOpen
	\bibfield  {author} {\bibinfo {author} {\bibfnamefont {T.}~\bibnamefont
			{{Clifton}}}, \bibinfo {author} {\bibfnamefont {P.~G.}\ \bibnamefont
			{{Ferreira}}}, \bibinfo {author} {\bibfnamefont {A.}~\bibnamefont
			{{Padilla}}}, \ and\ \bibinfo {author} {\bibfnamefont {C.}~\bibnamefont
			{{Skordis}}},\ }\href {\doibase 10.1016/j.physrep.2012.01.001} {\bibfield
		{journal} {\bibinfo  {journal} {\physrep}\ }\textbf {\bibinfo {volume}
			{513}},\ \bibinfo {pages} {1} (\bibinfo {year} {2012})},\ \Eprint
	{http://arxiv.org/abs/1106.2476} {arXiv:1106.2476 [astro-ph.CO]} \BibitemShut
	{NoStop}%
	\bibitem [{\citenamefont {{Nojiri}}\ and\ \citenamefont
		{{Odintsov}}(2011)}]{nojiri2011unified}%
	\BibitemOpen
	\bibfield  {author} {\bibinfo {author} {\bibfnamefont {S.}~\bibnamefont
			{{Nojiri}}}\ and\ \bibinfo {author} {\bibfnamefont {S.~D.}\ \bibnamefont
			{{Odintsov}}},\ }\href {\doibase 10.1016/j.physrep.2011.04.001} {\bibfield
		{journal} {\bibinfo  {journal} {\physrep}\ }\textbf {\bibinfo {volume}
			{505}},\ \bibinfo {pages} {59} (\bibinfo {year} {2011})},\ \Eprint
	{http://arxiv.org/abs/1011.0544} {arXiv:1011.0544 [gr-qc]} \BibitemShut
	{NoStop}%
	\bibitem [{\citenamefont {{Nordstr{\"o}m}}(1918)}]{nordstrom1918energy}%
	\BibitemOpen
	\bibfield  {author} {\bibinfo {author} {\bibfnamefont {G.}~\bibnamefont
			{{Nordstr{\"o}m}}},\ }\href@noop {} {\bibfield  {journal} {\bibinfo
			{journal} {Koninklijke Nederlandse Akademie van Wetenschappen Proceedings
				Series B Physical Sciences}\ }\textbf {\bibinfo {volume} {20}},\ \bibinfo
		{pages} {1238} (\bibinfo {year} {1918})}\BibitemShut {NoStop}%
	\bibitem [{\citenamefont {{Cai}}(2002)}]{cai2002gauss}%
	\BibitemOpen
	\bibfield  {author} {\bibinfo {author} {\bibfnamefont {R.-G.}\ \bibnamefont
			{{Cai}}},\ }\href {\doibase 10.1103/PhysRevD.65.084014} {\bibfield  {journal}
		{\bibinfo  {journal} {\prd}\ }\textbf {\bibinfo {volume} {65}},\ \bibinfo
		{eid} {084014} (\bibinfo {year} {2002})},\ \Eprint
	{http://arxiv.org/abs/hep-th/0109133} {arXiv:hep-th/0109133 [hep-th]}
	\BibitemShut {NoStop}%
	\bibitem [{\citenamefont {{Lovelock}}(1971)}]{lovelock1971einstein}%
	\BibitemOpen
	\bibfield  {author} {\bibinfo {author} {\bibfnamefont {D.}~\bibnamefont
			{{Lovelock}}},\ }\href {\doibase 10.1063/1.1665613} {\bibfield  {journal}
		{\bibinfo  {journal} {Journal of Mathematical Physics}\ }\textbf {\bibinfo
			{volume} {12}},\ \bibinfo {pages} {498} (\bibinfo {year} {1971})}\BibitemShut
	{NoStop}%
	\bibitem [{\citenamefont {{Dadhich}}\ \emph {et~al.}(2000)\citenamefont
		{{Dadhich}}, \citenamefont {{Maartens}}, \citenamefont {{Papadopoulos}},\
		and\ \citenamefont {{Rezania}}}]{dadhich2000black}%
	\BibitemOpen
	\bibfield  {author} {\bibinfo {author} {\bibfnamefont {N.}~\bibnamefont
			{{Dadhich}}}, \bibinfo {author} {\bibfnamefont {R.}~\bibnamefont
			{{Maartens}}}, \bibinfo {author} {\bibfnamefont {P.}~\bibnamefont
			{{Papadopoulos}}}, \ and\ \bibinfo {author} {\bibfnamefont {V.}~\bibnamefont
			{{Rezania}}},\ }\href {\doibase 10.1016/S0370-2693(00)00798-X} {\bibfield
		{journal} {\bibinfo  {journal} {Physics Letters B}\ }\textbf {\bibinfo
			{volume} {487}},\ \bibinfo {pages} {1} (\bibinfo {year} {2000})},\ \Eprint
	{http://arxiv.org/abs/hep-th/0003061} {arXiv:hep-th/0003061 [hep-th]}
	\BibitemShut {NoStop}%
	\bibitem [{\citenamefont {{Kirilin}}\ and\ \citenamefont
		{{Khriplovich}}(2002)}]{kirilin2002quantum}%
	\BibitemOpen
	\bibfield  {author} {\bibinfo {author} {\bibfnamefont {G.~G.}\ \bibnamefont
			{{Kirilin}}}\ and\ \bibinfo {author} {\bibfnamefont {I.~B.}\ \bibnamefont
			{{Khriplovich}}},\ }\href {\doibase 10.1134/1.1537290} {\bibfield  {journal}
		{\bibinfo  {journal} {Soviet Journal of Experimental and Theoretical
				Physics}\ }\textbf {\bibinfo {volume} {95}},\ \bibinfo {pages} {981}
		(\bibinfo {year} {2002})},\ \Eprint {http://arxiv.org/abs/gr-qc/0207118}
	{arXiv:gr-qc/0207118 [gr-qc]} \BibitemShut {NoStop}%
	\bibitem [{\citenamefont {{Bjerrum-Bohr}}\ \emph {et~al.}(2003)\citenamefont
		{{Bjerrum-Bohr}}, \citenamefont {{Donoghue}},\ and\ \citenamefont
		{{Holstein}}}]{bjerrum2003quantum}%
	\BibitemOpen
	\bibfield  {author} {\bibinfo {author} {\bibfnamefont {N.~E.}\ \bibnamefont
			{{Bjerrum-Bohr}}}, \bibinfo {author} {\bibfnamefont {J.~F.}\ \bibnamefont
			{{Donoghue}}}, \ and\ \bibinfo {author} {\bibfnamefont {B.~R.}\ \bibnamefont
			{{Holstein}}},\ }\href {\doibase 10.1103/PhysRevD.67.084033} {\bibfield
		{journal} {\bibinfo  {journal} {\prd}\ }\textbf {\bibinfo {volume} {67}},\
		\bibinfo {eid} {084033} (\bibinfo {year} {2003})},\ \Eprint
	{http://arxiv.org/abs/hep-th/0211072} {arXiv:hep-th/0211072 [hep-th]}
	\BibitemShut {NoStop}%
	\bibitem [{\citenamefont {{Casadio}}\ \emph {et~al.}(2002)\citenamefont
		{{Casadio}}, \citenamefont {{Fabbri}},\ and\ \citenamefont
		{{Mazzacurati}}}]{casadio2002new}%
	\BibitemOpen
	\bibfield  {author} {\bibinfo {author} {\bibfnamefont {R.}~\bibnamefont
			{{Casadio}}}, \bibinfo {author} {\bibfnamefont {A.}~\bibnamefont {{Fabbri}}},
		\ and\ \bibinfo {author} {\bibfnamefont {L.}~\bibnamefont {{Mazzacurati}}},\
	}\href {\doibase 10.1103/PhysRevD.65.084040} {\bibfield  {journal} {\bibinfo
			{journal} {\prd}\ }\textbf {\bibinfo {volume} {65}},\ \bibinfo {eid} {084040}
		(\bibinfo {year} {2002})},\ \Eprint {http://arxiv.org/abs/gr-qc/0111072}
	{arXiv:gr-qc/0111072 [gr-qc]} \BibitemShut {NoStop}%
	\bibitem [{\citenamefont {Devi}\ \emph {et~al.}(2023)\citenamefont {Devi},
		\citenamefont {Seenivasan}, \citenamefont {Chakrabarti},\ and\ \citenamefont
		{Majhi}}]{Devi:2021ctm}%
	\BibitemOpen
	\bibfield  {author} {\bibinfo {author} {\bibfnamefont {S.}~\bibnamefont
			{Devi}}, \bibinfo {author} {\bibfnamefont {A.~N.}\ \bibnamefont
			{Seenivasan}}, \bibinfo {author} {\bibfnamefont {S.}~\bibnamefont
			{Chakrabarti}}, \ and\ \bibinfo {author} {\bibfnamefont {B.~R.}\ \bibnamefont
			{Majhi}},\ }\href {\doibase 10.1016/j.dark.2023.101173} {\bibfield  {journal}
		{\bibinfo  {journal} {Phys. Dark Univ.}\ }\textbf {\bibinfo {volume} {39}},\
		\bibinfo {pages} {101173} (\bibinfo {year} {2023})},\ \Eprint
	{http://arxiv.org/abs/2105.11847} {arXiv:2105.11847 [gr-qc]} \BibitemShut
	{NoStop}%
	\bibitem [{\citenamefont
		{{Chandrasekhar}}(1998)}]{chandrasekhar1998mathematical}%
	\BibitemOpen
	\bibfield  {author} {\bibinfo {author} {\bibfnamefont {S.}~\bibnamefont
			{{Chandrasekhar}}},\ }\href@noop {} {\emph {\bibinfo {title} {{The
					Mathematical Theory of Black Holes}}}}\ (\bibinfo {year} {1998})\BibitemShut
	{NoStop}%
	\bibitem [{\citenamefont {{Cardoso}}\ and\ \citenamefont
		{{Pani}}(2019)}]{cardoso2019testing}%
	\BibitemOpen
	\bibfield  {author} {\bibinfo {author} {\bibfnamefont {V.}~\bibnamefont
			{{Cardoso}}}\ and\ \bibinfo {author} {\bibfnamefont {P.}~\bibnamefont
			{{Pani}}},\ }\href {\doibase 10.1007/s41114-019-0020-4} {\bibfield  {journal}
		{\bibinfo  {journal} {Living Reviews in Relativity}\ }\textbf {\bibinfo
			{volume} {22}},\ \bibinfo {eid} {4} (\bibinfo {year} {2019})},\ \Eprint
	{http://arxiv.org/abs/1904.05363} {arXiv:1904.05363 [gr-qc]} \BibitemShut
	{NoStop}%
	\bibitem [{\citenamefont {{Bardeen}}\ \emph {et~al.}(1973)\citenamefont
		{{Bardeen}}, \citenamefont {{Carter}},\ and\ \citenamefont
		{{Hawking}}}]{bardeen1973four}%
	\BibitemOpen
	\bibfield  {author} {\bibinfo {author} {\bibfnamefont {J.~M.}\ \bibnamefont
			{{Bardeen}}}, \bibinfo {author} {\bibfnamefont {B.}~\bibnamefont {{Carter}}},
		\ and\ \bibinfo {author} {\bibfnamefont {S.~W.}\ \bibnamefont {{Hawking}}},\
	}\href {\doibase 10.1007/BF01645742} {\bibfield  {journal} {\bibinfo
			{journal} {Communications in Mathematical Physics}\ }\textbf {\bibinfo
			{volume} {31}},\ \bibinfo {pages} {161} (\bibinfo {year} {1973})}\BibitemShut
	{NoStop}%
	\bibitem [{\citenamefont {{Mafuz}}\ \emph {et~al.}(2024)\citenamefont
		{{Mafuz}}, \citenamefont {{Diwan}}, \citenamefont {{Jana}},\ and\
		\citenamefont {{Kar}}}]{mafuz2024shadows}%
	\BibitemOpen
	\bibfield  {author} {\bibinfo {author} {\bibfnamefont {M.~G.}\ \bibnamefont
			{{Mafuz}}}, \bibinfo {author} {\bibfnamefont {R.}~\bibnamefont {{Diwan}}},
		\bibinfo {author} {\bibfnamefont {S.}~\bibnamefont {{Jana}}}, \ and\ \bibinfo
		{author} {\bibfnamefont {S.}~\bibnamefont {{Kar}}},\ }\href {\doibase
		10.1140/epjp/s13360-024-04993-8} {\bibfield  {journal} {\bibinfo  {journal}
			{European Physical Journal Plus}\ }\textbf {\bibinfo {volume} {139}},\
		\bibinfo {eid} {219} (\bibinfo {year} {2024})},\ \Eprint
	{http://arxiv.org/abs/2309.11383} {arXiv:2309.11383 [gr-qc]} \BibitemShut
	{NoStop}%
	\bibitem [{\citenamefont {{Chattopadhyay}}\ and\ \citenamefont
		{{Ryu}}(2009)}]{chatto2009effects}%
	\BibitemOpen
	\bibfield  {author} {\bibinfo {author} {\bibfnamefont {I.}~\bibnamefont
			{{Chattopadhyay}}}\ and\ \bibinfo {author} {\bibfnamefont {D.}~\bibnamefont
			{{Ryu}}},\ }\href {\doibase 10.1088/0004-637X/694/1/492} {\bibfield
		{journal} {\bibinfo  {journal} {\apj}\ }\textbf {\bibinfo {volume} {694}},\
		\bibinfo {pages} {492} (\bibinfo {year} {2009})},\ \Eprint
	{http://arxiv.org/abs/0812.2607} {arXiv:0812.2607 [astro-ph]} \BibitemShut
	{NoStop}%
	\bibitem [{\citenamefont {{Kiselev}}(2003)}]{kiselev2003quintessence}%
	\BibitemOpen
	\bibfield  {author} {\bibinfo {author} {\bibfnamefont {V.~V.}\ \bibnamefont
			{{Kiselev}}},\ }\href {\doibase 10.1088/0264-9381/20/6/310} {\bibfield
		{journal} {\bibinfo  {journal} {Classical and Quantum Gravity}\ }\textbf
		{\bibinfo {volume} {20}},\ \bibinfo {pages} {1187} (\bibinfo {year}
		{2003})},\ \Eprint {http://arxiv.org/abs/gr-qc/0210040} {arXiv:gr-qc/0210040
		[gr-qc]} \BibitemShut {NoStop}%
	\bibitem [{\citenamefont {{Heydarzade}}\ and\ \citenamefont
		{{Darabi}}(2017)}]{heydarzade2017black}%
	\BibitemOpen
	\bibfield  {author} {\bibinfo {author} {\bibfnamefont {Y.}~\bibnamefont
			{{Heydarzade}}}\ and\ \bibinfo {author} {\bibfnamefont {F.}~\bibnamefont
			{{Darabi}}},\ }\href {\doibase 10.1016/j.physletb.2017.05.064} {\bibfield
		{journal} {\bibinfo  {journal} {Physics Letters B}\ }\textbf {\bibinfo
			{volume} {771}},\ \bibinfo {pages} {365} (\bibinfo {year} {2017})},\ \Eprint
	{http://arxiv.org/abs/1702.07766} {arXiv:1702.07766 [gr-qc]} \BibitemShut
	{NoStop}%
	\bibitem [{\citenamefont {{Ca{\~n}ate}}\ and\ \citenamefont
		{{Bergliaffa}}(2020)}]{canate2020novel}%
	\BibitemOpen
	\bibfield  {author} {\bibinfo {author} {\bibfnamefont {P.}~\bibnamefont
			{{Ca{\~n}ate}}}\ and\ \bibinfo {author} {\bibfnamefont {S.~E.~P.}\
			\bibnamefont {{Bergliaffa}}},\ }\href {\doibase 10.1103/PhysRevD.102.104038}
	{\bibfield  {journal} {\bibinfo  {journal} {\prd}\ }\textbf {\bibinfo
			{volume} {102}},\ \bibinfo {eid} {104038} (\bibinfo {year} {2020})},\ \Eprint
	{http://arxiv.org/abs/2010.04858} {arXiv:2010.04858 [gr-qc]} \BibitemShut
	{NoStop}%
	\bibitem [{\citenamefont {{Chakrabarti}}(1989)}]{chakrabarti1989standing}%
	\BibitemOpen
	\bibfield  {author} {\bibinfo {author} {\bibfnamefont {S.~K.}\ \bibnamefont
			{{Chakrabarti}}},\ }\href {\doibase 10.1086/168125} {\bibfield  {journal}
		{\bibinfo  {journal} {\apj}\ }\textbf {\bibinfo {volume} {347}},\ \bibinfo
		{pages} {365} (\bibinfo {year} {1989})}\BibitemShut {NoStop}%
	\bibitem [{\citenamefont {{Das}}\ \emph {et~al.}(2001)\citenamefont {{Das}},
		\citenamefont {{Chattopadhyay}},\ and\ \citenamefont
		{{Chakrabarti}}}]{das2001standing}%
	\BibitemOpen
	\bibfield  {author} {\bibinfo {author} {\bibfnamefont {S.}~\bibnamefont
			{{Das}}}, \bibinfo {author} {\bibfnamefont {I.}~\bibnamefont
			{{Chattopadhyay}}}, \ and\ \bibinfo {author} {\bibfnamefont {S.~K.}\
			\bibnamefont {{Chakrabarti}}},\ }\href {\doibase 10.1086/321692} {\bibfield
		{journal} {\bibinfo  {journal} {\apj}\ }\textbf {\bibinfo {volume} {557}},\
		\bibinfo {pages} {983} (\bibinfo {year} {2001})},\ \Eprint
	{http://arxiv.org/abs/astro-ph/0107046} {arXiv:astro-ph/0107046 [astro-ph]}
	\BibitemShut {NoStop}%
	\bibitem [{\citenamefont {{Dihingia}}\ \emph
		{et~al.}(2019{\natexlab{a}})\citenamefont {{Dihingia}}, \citenamefont
		{{Das}}, \citenamefont {{Maity}},\ and\ \citenamefont
		{{Nandi}}}]{dihingia2019shocks}%
	\BibitemOpen
	\bibfield  {author} {\bibinfo {author} {\bibfnamefont {I.~K.}\ \bibnamefont
			{{Dihingia}}}, \bibinfo {author} {\bibfnamefont {S.}~\bibnamefont {{Das}}},
		\bibinfo {author} {\bibfnamefont {D.}~\bibnamefont {{Maity}}}, \ and\
		\bibinfo {author} {\bibfnamefont {A.}~\bibnamefont {{Nandi}}},\ }\href
	{\doibase 10.1093/mnras/stz1933} {\bibfield  {journal} {\bibinfo  {journal}
			{\mnras}\ }\textbf {\bibinfo {volume} {488}},\ \bibinfo {pages} {2412}
		(\bibinfo {year} {2019}{\natexlab{a}})},\ \Eprint
	{http://arxiv.org/abs/1903.02856} {arXiv:1903.02856 [astro-ph.HE]}
	\BibitemShut {NoStop}%
	\bibitem [{\citenamefont {{Dihingia}}\ \emph {et~al.}(2018)\citenamefont
		{{Dihingia}}, \citenamefont {{Das}}, \citenamefont {{Maity}},\ and\
		\citenamefont {{Chakrabarti}}}]{dihingia2018limitations}%
	\BibitemOpen
	\bibfield  {author} {\bibinfo {author} {\bibfnamefont {I.~K.}\ \bibnamefont
			{{Dihingia}}}, \bibinfo {author} {\bibfnamefont {S.}~\bibnamefont {{Das}}},
		\bibinfo {author} {\bibfnamefont {D.}~\bibnamefont {{Maity}}}, \ and\
		\bibinfo {author} {\bibfnamefont {S.}~\bibnamefont {{Chakrabarti}}},\ }\href
	{\doibase 10.1103/PhysRevD.98.083004} {\bibfield  {journal} {\bibinfo
			{journal} {\prd}\ }\textbf {\bibinfo {volume} {98}},\ \bibinfo {eid} {083004}
		(\bibinfo {year} {2018})},\ \Eprint {http://arxiv.org/abs/1806.08481}
	{arXiv:1806.08481 [astro-ph.HE]} \BibitemShut {NoStop}%
	\bibitem [{\citenamefont {{Riffert}}\ and\ \citenamefont
		{{Herold}}(1995)}]{riffert1995relativistic}%
	\BibitemOpen
	\bibfield  {author} {\bibinfo {author} {\bibfnamefont {H.}~\bibnamefont
			{{Riffert}}}\ and\ \bibinfo {author} {\bibfnamefont {H.}~\bibnamefont
			{{Herold}}},\ }\href {\doibase 10.1086/176161} {\bibfield  {journal}
		{\bibinfo  {journal} {\apj}\ }\textbf {\bibinfo {volume} {450}},\ \bibinfo
		{pages} {508} (\bibinfo {year} {1995})}\BibitemShut {NoStop}%
	\bibitem [{\citenamefont {{Peitz}}\ and\ \citenamefont
		{{Appl}}(1997)}]{peitz1997viscous}%
	\BibitemOpen
	\bibfield  {author} {\bibinfo {author} {\bibfnamefont {J.}~\bibnamefont
			{{Peitz}}}\ and\ \bibinfo {author} {\bibfnamefont {S.}~\bibnamefont
			{{Appl}}},\ }\bibfield  {booktitle} {\emph {\bibinfo {booktitle} {Accretion
				Disks - New Aspects}},\ }\href {\doibase 10.1007/BFb0105834} {\ \textbf
		{\bibinfo {volume} {487}},\ \bibinfo {pages} {209} (\bibinfo {year}
		{1997})}\BibitemShut {NoStop}%
	\bibitem [{\citenamefont {{Dihingia}}\ \emph
		{et~al.}(2019{\natexlab{b}})\citenamefont {{Dihingia}}, \citenamefont
		{{Das}},\ and\ \citenamefont {{Nandi}}}]{dihingia2019low}%
	\BibitemOpen
	\bibfield  {author} {\bibinfo {author} {\bibfnamefont {I.~K.}\ \bibnamefont
			{{Dihingia}}}, \bibinfo {author} {\bibfnamefont {S.}~\bibnamefont {{Das}}}, \
		and\ \bibinfo {author} {\bibfnamefont {A.}~\bibnamefont {{Nandi}}},\ }\href
	{\doibase 10.1093/mnras/stz168} {\bibfield  {journal} {\bibinfo  {journal}
			{\mnras}\ }\textbf {\bibinfo {volume} {484}},\ \bibinfo {pages} {3209}
		(\bibinfo {year} {2019}{\natexlab{b}})},\ \Eprint
	{http://arxiv.org/abs/1901.04293} {arXiv:1901.04293 [astro-ph.HE]}
	\BibitemShut {NoStop}%
	\bibitem [{\citenamefont {{Sen}}\ \emph {et~al.}(2022)\citenamefont {{Sen}},
		\citenamefont {{Maity}},\ and\ \citenamefont {{Das}}}]{Sen-etal2022}%
	\BibitemOpen
	\bibfield  {author} {\bibinfo {author} {\bibfnamefont {G.}~\bibnamefont
			{{Sen}}}, \bibinfo {author} {\bibfnamefont {D.}~\bibnamefont {{Maity}}}, \
		and\ \bibinfo {author} {\bibfnamefont {S.}~\bibnamefont {{Das}}},\ }\href
	{\doibase 10.1088/1475-7516/2022/08/048} {\bibfield  {journal} {\bibinfo
			{journal} {\jcap}\ }\textbf {\bibinfo {volume} {2022}},\ \bibinfo {eid} {048}
		(\bibinfo {year} {2022})},\ \Eprint {http://arxiv.org/abs/2204.02110}
	{arXiv:2204.02110 [astro-ph.HE]} \BibitemShut {NoStop}%
	\bibitem [{\citenamefont {{Chattopadhyay}}\ and\ \citenamefont
		{{Kumar}}(2016)}]{chatto2016estimation}%
	\BibitemOpen
	\bibfield  {author} {\bibinfo {author} {\bibfnamefont {I.}~\bibnamefont
			{{Chattopadhyay}}}\ and\ \bibinfo {author} {\bibfnamefont {R.}~\bibnamefont
			{{Kumar}}},\ }\href {\doibase 10.1093/mnras/stw876} {\bibfield  {journal}
		{\bibinfo  {journal} {\mnras}\ }\textbf {\bibinfo {volume} {459}},\ \bibinfo
		{pages} {3792} (\bibinfo {year} {2016})},\ \Eprint
	{http://arxiv.org/abs/1605.00752} {arXiv:1605.00752 [astro-ph.HE]}
	\BibitemShut {NoStop}%
	\bibitem [{\citenamefont {{Dihingia}}\ \emph {et~al.}(2020)\citenamefont
		{{Dihingia}}, \citenamefont {{Maity}}, \citenamefont {{Chakrabarti}},\ and\
		\citenamefont {{Das}}}]{dihingia2020study}%
	\BibitemOpen
	\bibfield  {author} {\bibinfo {author} {\bibfnamefont {I.~K.}\ \bibnamefont
			{{Dihingia}}}, \bibinfo {author} {\bibfnamefont {D.}~\bibnamefont {{Maity}}},
		\bibinfo {author} {\bibfnamefont {S.}~\bibnamefont {{Chakrabarti}}}, \ and\
		\bibinfo {author} {\bibfnamefont {S.}~\bibnamefont {{Das}}},\ }\href
	{\doibase 10.1103/PhysRevD.102.023012} {\bibfield  {journal} {\bibinfo
			{journal} {\prd}\ }\textbf {\bibinfo {volume} {102}},\ \bibinfo {eid}
		{023012} (\bibinfo {year} {2020})},\ \Eprint
	{http://arxiv.org/abs/2004.03195} {arXiv:2004.03195 [astro-ph.HE]}
	\BibitemShut {NoStop}%
	\bibitem [{\citenamefont {{Liang}}\ and\ \citenamefont
		{{Thompson}}(1980)}]{liang1980transonic}%
	\BibitemOpen
	\bibfield  {author} {\bibinfo {author} {\bibfnamefont {E.~P.~T.}\
			\bibnamefont {{Liang}}}\ and\ \bibinfo {author} {\bibfnamefont {K.~A.}\
			\bibnamefont {{Thompson}}},\ }\href {\doibase 10.1086/158231} {\bibfield
		{journal} {\bibinfo  {journal} {\apj}\ }\textbf {\bibinfo {volume} {240}},\
		\bibinfo {pages} {271} (\bibinfo {year} {1980})}\BibitemShut {NoStop}%
	\bibitem [{\citenamefont {{Das}}(2007)}]{das2007behaviour}%
	\BibitemOpen
	\bibfield  {author} {\bibinfo {author} {\bibfnamefont {S.}~\bibnamefont
			{{Das}}},\ }\href {\doibase 10.1111/j.1365-2966.2007.11501.x} {\bibfield
		{journal} {\bibinfo  {journal} {\mnras}\ }\textbf {\bibinfo {volume} {376}},\
		\bibinfo {pages} {1659} (\bibinfo {year} {2007})},\ \Eprint
	{http://arxiv.org/abs/astro-ph/0610651} {arXiv:astro-ph/0610651 [astro-ph]}
	\BibitemShut {NoStop}%
	\bibitem [{\citenamefont {{Kumar}}\ \emph {et~al.}(2013)\citenamefont
		{{Kumar}}, \citenamefont {{Singh}}, \citenamefont {{Chattopadhyay}},\ and\
		\citenamefont {{Chakrabarti}}}]{kumar2013effect}%
	\BibitemOpen
	\bibfield  {author} {\bibinfo {author} {\bibfnamefont {R.}~\bibnamefont
			{{Kumar}}}, \bibinfo {author} {\bibfnamefont {C.~B.}\ \bibnamefont
			{{Singh}}}, \bibinfo {author} {\bibfnamefont {I.}~\bibnamefont
			{{Chattopadhyay}}}, \ and\ \bibinfo {author} {\bibfnamefont {S.~K.}\
			\bibnamefont {{Chakrabarti}}},\ }\href {\doibase 10.1093/mnras/stt1781}
	{\bibfield  {journal} {\bibinfo  {journal} {\mnras}\ }\textbf {\bibinfo
			{volume} {436}},\ \bibinfo {pages} {2864} (\bibinfo {year} {2013})},\ \Eprint
	{http://arxiv.org/abs/1310.0144} {arXiv:1310.0144 [astro-ph.HE]} \BibitemShut
	{NoStop}%
	\bibitem [{\citenamefont {{Kato}}\ \emph {et~al.}(1993)\citenamefont {{Kato}},
		\citenamefont {{Wu}}, \citenamefont {{Yang}},\ and\ \citenamefont
		{{Yang}}}]{kato1993sonic}%
	\BibitemOpen
	\bibfield  {author} {\bibinfo {author} {\bibfnamefont {S.}~\bibnamefont
			{{Kato}}}, \bibinfo {author} {\bibfnamefont {X.-B.}\ \bibnamefont {{Wu}}},
		\bibinfo {author} {\bibfnamefont {L.-T.}\ \bibnamefont {{Yang}}}, \ and\
		\bibinfo {author} {\bibfnamefont {Z.-L.}\ \bibnamefont {{Yang}}},\ }\href
	{\doibase 10.1093/mnras/260.2.317} {\bibfield  {journal} {\bibinfo  {journal}
			{\mnras}\ }\textbf {\bibinfo {volume} {260}},\ \bibinfo {pages} {317}
		(\bibinfo {year} {1993})}\BibitemShut {NoStop}%
	\bibitem [{\citenamefont {{Chakrabarti}}\ and\ \citenamefont
		{{Das}}(2004)}]{Das2004shock}%
	\BibitemOpen
	\bibfield  {author} {\bibinfo {author} {\bibfnamefont {S.~K.}\ \bibnamefont
			{{Chakrabarti}}}\ and\ \bibinfo {author} {\bibfnamefont {S.}~\bibnamefont
			{{Das}}},\ }\href {\doibase 10.1111/j.1365-2966.2004.07536.x} {\bibfield
		{journal} {\bibinfo  {journal} {\mnras}\ }\textbf {\bibinfo {volume} {349}},\
		\bibinfo {pages} {649} (\bibinfo {year} {2004})},\ \Eprint
	{http://arxiv.org/abs/astro-ph/0402561} {arXiv:astro-ph/0402561 [astro-ph]}
	\BibitemShut {NoStop}%
	\bibitem [{\citenamefont {{Mitra}}\ and\ \citenamefont
		{{Das}}(2024)}]{mitra2024mhdshock}%
	\BibitemOpen
	\bibfield  {author} {\bibinfo {author} {\bibfnamefont {S.}~\bibnamefont
			{{Mitra}}}\ and\ \bibinfo {author} {\bibfnamefont {S.}~\bibnamefont
			{{Das}}},\ }\href {\doibase 10.3847/1538-4357/ad55cb} {\bibfield  {journal}
		{\bibinfo  {journal} {\apj}\ }\textbf {\bibinfo {volume} {971}},\ \bibinfo
		{eid} {28} (\bibinfo {year} {2024})},\ \Eprint
	{http://arxiv.org/abs/2405.16326} {arXiv:2405.16326 [astro-ph.HE]}
	\BibitemShut {NoStop}%
	\bibitem [{\citenamefont {{Quenby}}(2010)}]{quenby2010foundations}%
	\BibitemOpen
	\bibfield  {author} {\bibinfo {author} {\bibfnamefont {J.}~\bibnamefont
			{{Quenby}}},\ }\href {\doibase 10.1080/00107510903303667} {\enquote {\bibinfo
			{title} {{Foundations of High-Energy Astrophysics, by M. Vietri}},}\ }
	(\bibinfo {year} {2010})\BibitemShut {NoStop}%
	\bibitem [{\citenamefont {{Okuda}}\ \emph {et~al.}(2019)\citenamefont
		{{Okuda}}, \citenamefont {{Singh}}, \citenamefont {{Das}}, \citenamefont
		{{Aktar}}, \citenamefont {{Nandi}},\ and\ \citenamefont {{Dal
				Pino}}}]{okuda2019possible}%
	\BibitemOpen
	\bibfield  {author} {\bibinfo {author} {\bibfnamefont {T.}~\bibnamefont
			{{Okuda}}}, \bibinfo {author} {\bibfnamefont {C.~B.}\ \bibnamefont
			{{Singh}}}, \bibinfo {author} {\bibfnamefont {S.}~\bibnamefont {{Das}}},
		\bibinfo {author} {\bibfnamefont {R.}~\bibnamefont {{Aktar}}}, \bibinfo
		{author} {\bibfnamefont {A.}~\bibnamefont {{Nandi}}}, \ and\ \bibinfo
		{author} {\bibfnamefont {E.~M. d.~G.}\ \bibnamefont {{Dal Pino}}},\ }\href
	{\doibase 10.1093/pasj/psz021} {\bibfield  {journal} {\bibinfo  {journal}
			{\pasj}\ }\textbf {\bibinfo {volume} {71}},\ \bibinfo {eid} {49} (\bibinfo
		{year} {2019})},\ \Eprint {http://arxiv.org/abs/1902.02933} {arXiv:1902.02933
		[astro-ph.HE]} \BibitemShut {NoStop}%
	\bibitem [{\citenamefont {{Novikov}}\ and\ \citenamefont
		{{Thorne}}(1973)}]{novikov1973}%
	\BibitemOpen
	\bibfield  {author} {\bibinfo {author} {\bibfnamefont {I.~D.}\ \bibnamefont
			{{Novikov}}}\ and\ \bibinfo {author} {\bibfnamefont {K.~S.}\ \bibnamefont
			{{Thorne}}},\ }\bibfield  {booktitle} {\emph {\bibinfo {booktitle} {Black
				Holes (Les Astres Occlus)}},\ }\href@noop {} {\ ,\ \bibinfo {pages} {343}
		(\bibinfo {year} {1973})}\BibitemShut {NoStop}%
	\bibitem [{\citenamefont {{Karzas}}\ and\ \citenamefont
		{{Latter}}(1961)}]{karzas1961electron}%
	\BibitemOpen
	\bibfield  {author} {\bibinfo {author} {\bibfnamefont {W.~J.}\ \bibnamefont
			{{Karzas}}}\ and\ \bibinfo {author} {\bibfnamefont {R.}~\bibnamefont
			{{Latter}}},\ }\href {\doibase 10.1086/190063} {\bibfield  {journal}
		{\bibinfo  {journal} {\apjs}\ }\textbf {\bibinfo {volume} {6}},\ \bibinfo
		{pages} {167} (\bibinfo {year} {1961})}\BibitemShut {NoStop}%
	\bibitem [{\citenamefont {{Chattopadhyay}}\ and\ \citenamefont
		{{Chakrabarti}}(2002)}]{chattopadhyay2002radiatively}%
	\BibitemOpen
	\bibfield  {author} {\bibinfo {author} {\bibfnamefont {I.}~\bibnamefont
			{{Chattopadhyay}}}\ and\ \bibinfo {author} {\bibfnamefont {S.~K.}\
			\bibnamefont {{Chakrabarti}}},\ }\href {\doibase
		10.1046/j.1365-8711.2002.05424.x} {\bibfield  {journal} {\bibinfo  {journal}
			{\mnras}\ }\textbf {\bibinfo {volume} {333}},\ \bibinfo {pages} {454}
		(\bibinfo {year} {2002})},\ \Eprint {http://arxiv.org/abs/astro-ph/0202351}
	{arXiv:astro-ph/0202351 [astro-ph]} \BibitemShut {NoStop}%
	\bibitem [{\citenamefont {{Sen}}\ \emph {et~al.}(2024)\citenamefont {{Sen}},
		\citenamefont {{Maity}},\ and\ \citenamefont {{Das}}}]{Sen-etal2024}%
	\BibitemOpen
	\bibfield  {author} {\bibinfo {author} {\bibfnamefont {G.}~\bibnamefont
			{{Sen}}}, \bibinfo {author} {\bibfnamefont {D.}~\bibnamefont {{Maity}}}, \
		and\ \bibinfo {author} {\bibfnamefont {S.}~\bibnamefont {{Das}}},\ }\href
	{\doibase 10.1088/1475-7516/2024/10/068} {\bibfield  {journal} {\bibinfo
			{journal} {\jcap}\ }\textbf {\bibinfo {volume} {2024}},\ \bibinfo {eid} {068}
		(\bibinfo {year} {2024})},\ \Eprint {http://arxiv.org/abs/2405.11453}
	{arXiv:2405.11453 [astro-ph.HE]} \BibitemShut {NoStop}%
	\bibitem [{\citenamefont {{Patra}}\ \emph {et~al.}(2024)\citenamefont
		{{Patra}}, \citenamefont {{Majhi}},\ and\ \citenamefont
		{{Das}}}]{Patra-etal2024}%
	\BibitemOpen
	\bibfield  {author} {\bibinfo {author} {\bibfnamefont {S.}~\bibnamefont
			{{Patra}}}, \bibinfo {author} {\bibfnamefont {B.~R.}\ \bibnamefont
			{{Majhi}}}, \ and\ \bibinfo {author} {\bibfnamefont {S.}~\bibnamefont
			{{Das}}},\ }\href {\doibase 10.1088/1475-7516/2024/01/060} {\bibfield
		{journal} {\bibinfo  {journal} {\jcap}\ }\textbf {\bibinfo {volume} {2024}},\
		\bibinfo {eid} {060} (\bibinfo {year} {2024})},\ \Eprint
	{http://arxiv.org/abs/2308.12839} {arXiv:2308.12839 [astro-ph.HE]}
	\BibitemShut {NoStop}%
	\bibitem [{\citenamefont {{Abramowitz}}\ and\ \citenamefont
		{{Stegun}}(1965)}]{abramowitz1965handbook}%
	\BibitemOpen
	\bibfield  {author} {\bibinfo {author} {\bibfnamefont {M.}~\bibnamefont
			{{Abramowitz}}}\ and\ \bibinfo {author} {\bibfnamefont {I.~A.}\ \bibnamefont
			{{Stegun}}},\ }\href@noop {} {\emph {\bibinfo {title} {{Handbook of
					mathematical functions with formulas, graphs, and mathematical tables}}}}\
	(\bibinfo {year} {1965})\BibitemShut {NoStop}%
\end{thebibliography}
%

\end{document}